\documentclass[aps,onecolumn,preprint,superscriptaddress,prl,showpacs,preprintnumbers]{revtex4-1}
\usepackage{mathrsfs}
\usepackage{amsmath}
\usepackage{amsfonts}
\usepackage{amssymb}
\usepackage{amsthm}
\usepackage{graphicx}
\usepackage{natbib}
\usepackage{color}
\usepackage{hyperref}
\usepackage{bm}
\usepackage[caption=false]{subfig}
\usepackage{verbatim}
\setcitestyle{super}

\newcommand{\vct}[1]{\boldsymbol{#1}}
\newcommand{\uvv}[1]{\hat{\boldsymbol{#1}}}
\newcommand{\m}{\vct{m}}
\newcommand{\Hpar}{H_\parallel}
\newcommand{\Hper}{H_\perp}
\newcommand{\Hext}{\vct{H}_\text{ext}}
\newcommand{\Aex}{A_{\text{ex}}}
\newcommand{\Is}{\vct{J}^s}

\newcommand{\Gp}{G_{\perp}}
\newcommand{\GI}{G_{\text{I}}}
\newcommand{\GN}{G_{\text{N}}}
\renewcommand{\th}{d}
\newcommand{\Tau}{\mathrm{T}}

\begin{document}

\title{Proposal for a Domain Wall Nano-Oscillator driven by Non-uniform Spin Currents}

\author{Sanchar Sharma}
\affiliation{Department of Electrical Engineering, Indian Institute of Technology Bombay, Powai, Mumbai-400076, India}

\author{Bhaskaran Muralidharan*}
\email{bm@ee.iitb.ac.in}
\affiliation{Department of Electrical Engineering, Indian Institute of Technology Bombay, Powai, Mumbai-400076, India}

\author{Ashwin Tulapurkar}
\affiliation{Department of Electrical Engineering, Indian Institute of Technology Bombay, Powai, Mumbai-400076, India}



\begin{abstract}

We propose a new mechanism and a related device concept for a robust, magnetic field tunable radio-frequency (rf) oscillator using the self oscillation of a magnetic domain wall subject to a uniform static magnetic field and a spatially non-uniform vertical dc spin current. The self oscillation of the domain wall is created as it translates periodically between two unstable positions, one being in the region where both the dc spin current and the magnetic field are present, and the other, being where only the magnetic field is present. The vertical dc spin current pushes it away from one unstable position while the magnetic field pushes it away from the other. We show that such oscillations are stable under noise and can exhibit a quality factor of over 1000.  A domain wall under dynamic translation, not only being a source for rich physics, is also a promising candidate for advancements in nanoelectronics with the actively researched racetrack memory architecture, digital and analog switching paradigms as candidate examples. Devising a stable rf oscillator using a domain wall is hence another step towards the realization of an all domain wall logic scheme.

\end{abstract}

\flushbottom
\maketitle

\thispagestyle{empty}

Self oscillators show a remarkable property of sustaining oscillatory behavior without being driven by sources that possess inherent periodicity. In the macroscopic world, a few well known examples of self oscillations include the heartbeat, violin string oscillations in response to steady bowing \cite{Music}, the Vander Pol oscillator and the infamous collapse of the Tacoma Narrows Bridge in 1940\cite{Tacoma}. In the nanoscale too, self oscillations govern the underlying principle of the well studied resonant tunneling diode based oscillator, the spin torque oscillator \cite{kise,Kim_STO,Grollier} and also, the recently noted phenomenon of nuclear spin induced current oscillations in quantum dots \cite{Ono}. Our proposal relies on the self oscillations in the translatory motion of a magnetic domain wall.

The interest in dynamics of magnetic domain walls has been active for decades\cite{QCoh,Soliton,MagDW,MagDW2,MagNoise,PinMagDW}, recently intensified by the discovery of current-driven domain wall motion\cite{CurDWExp,CurDWExp2,CurDWTh} and its related applications in nanoelectronics\cite{Rev_Mod_DW,RaceTrack,Grollier,ExpDW}. Field driven oscillations have also been observed and studied \cite{MagOsc,MagNoise} for a long time. However, these oscillations are accompanied by a drift which makes them unusable as a device. On a different note, vertical injection of uniform spin current is proposed as a means for high domain wall velocities \cite{VerCurExp1,VerCurExp,VerCurTh}. Here, we propose stable oscillations caused by a constant magnetic field whose drift is canceled by a vertically injected non-uniform spin current as depicted in the schematic in Fig.~1(a), thereby resulting in a stable periodic motion. 

Normally, domain wall motion under vertical spin currents is caused by {\it{field like torque}} which is typically smaller than the {\it{Slonczewski like torque}}\cite{Slon,VerCurTh,MechSTT,STDiode} usually responsible for the switching of the free layer of magnetic tunnel junctions (MTJ) \cite{Ralph_STT}. However, the presence of a magnetic field allows an efficient transfer of the Slonczewski like torque without any need of a field like term, as shown later. We note that current-driven domain wall oscillations with a drift have been observed\cite{CurOsc_Pos} while stable oscillations on a pinned domain wall were proposed\cite{StableOsc}. Our proposal differs from the latter in the control of oscillations via an external field instead of a pinning potential, both of which under rigid domain wall approximation work to rotate the domain wall. 
\begin{figure*}
	\includegraphics[width=\textwidth,keepaspectratio]{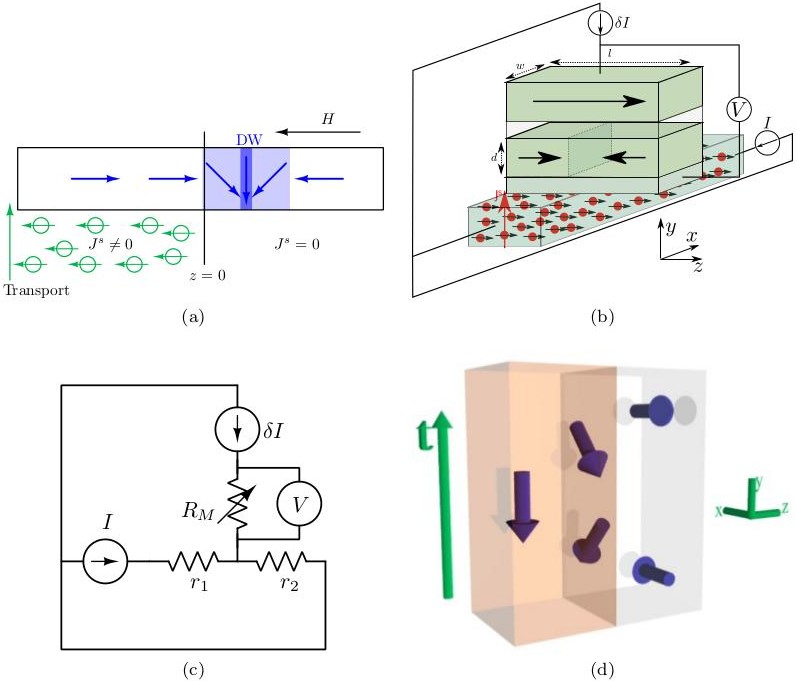}
\caption{Working principle and device design. (a) Schematic depicting the principal idea of the proposal which comprises a domain wall magnet with the incident non-uniform spin current and a uniform magnetic field. The domain wall is shown at an arbitrary instantaneous position. The non-uniform spin current is incident on a region spanning one-half of the magnet. (b) A 3D schematic of the proposed device design. A strip of Tantalum localized in the desired region of the domain wall magnet, connected to a current source generates the desired spin current. The applied magnetic field along the $-z$ axis is not shown here for clarity. (c) A circuit diagram representing the device schematic in (b). $r_1$ is the resistance of the part of Tantalum strip between the constant current source, `I', and MTJ, and similarly $r_2$. (d) Zoomed-in motion of middle spin marked in (a), depicting the rotation it undergoes which gets converted into motion via its hard axis anisotropy. The orange colored region is the portion where spin current is non-zero.} \label{Fig:Setup}
\end{figure*}

We first provide a theoretical analysis using a rigid domain wall model to find an approximate waveform for the oscillations. We show that the frequency is twice the resonant frequency of a magnet in a magnetic field; while the amplitude is approximately a linear function of the ratio of the hard axis anisotropy and the magnetic field. The oscillatory part of the waveform is independent of the input spin current to a very good degree of accuracy; and hence can be of great technological advantage for accurate oscillatory waveforms. We then also numerically simulate the micromagnetic Landau-Lifshitz-Gilbert equation (see supplementary material) as a verification of the oscillations and find the frequency to be in good agreement with the analytic result. Lastly, we analyze the effect of thermal noise on the oscillator and find that it is robust at room temperature by numerically calculating its Q-factor.


The working principle of the proposed oscillator is depicted in Fig.~1(a). The vertically incident spin current is spatially confined to the left half of the domain wall, while the uniform magnetic field along the $-z$ direction exists throughout the region. Both the regions, $z\gg0$ and $z\ll0$ by themselves are unstable for the domain wall (unless the spin current is too low) and hence it is restricted to be in some region around $z=0$. A possible device realization of the proposal, in which the domain wall self-oscillations may be effectively translated into an alternating current oscillations is depicted in Fig.~1(b). A dc-current source is connected to the bottom strip of Tantalum, which is used to inject a spin current via the giant spin hall effect (SHE)\cite{Khav,GSHE_Ta,Ssalahud}. The strip of Tantalum conveniently acts as a source of spin current that injects a constant spin current in a locally confined region marked in the schematic in Fig.~1(a). An MTJ like structure is then used to sense the position of the domain wall via a measurement of the change in resistance\cite{VerCurExp1}. A small current is applied and the corresponding voltage across the MTJ is measured. An equivalent circuit for the entire set up is shown in Fig.~1(c) depicting the measurement in a more explicit way. We assume that the current used for the measurement is small enough such that it doesn't have any additional effect on the domain wall dynamics. 

The analysis to follow will be based on Fig.~1(a) which captures the essence of our proposal. The region of non-zero spin current ``pushes'' the domain wall towards the region of zero spin current via spin transfer torque (STT). However, the domain wall cannot keep moving away from the spin current as the magnetic field will push it back via the dissipative Gilbert term\cite{MagDW}. As shown in Fig.~1(d), when the domain wall enters the spin current region, it is reflected back with a different azimuthal angle accumulated because of the magnetic field. The hard axis anisotropy then keeps the domain wall moving until the magnetic field rotates the domain wall again to cause reverse motion. Hence, the magnetic field causes a perpetual rotation, while the hard axis anisotropy converts the rotation into a translation of the domain wall. The spin current then acts as the energy input which negates the dissipative effect present in a purely field driven motion, hence stopping the drift observed in the latter case. For a typical case of a low dissipation constant, the domain wall will need only a small amount of push from the spin current and hence the oscillations will be almost independent of it. The average location of the domain wall though will be dependent on the spin current density.
Our analysis is based on the magnetization dynamics of the domain wall described by the Landau-Lifshitz-Gilbert equation augmented by the Slonczewski spin torque term\cite{Slon} given by
\begin{equation} \label{Eq:LLG}
	\frac{d \m}{dt} = -\gamma \m\times \left( \Hext + \vct{H}_{\text{eff}}\right) + \alpha\m\times\left(\frac{d\m}{dt}\right) - \frac{\gamma \hbar}{e\mu_0 M_s \th}\m\times\left(\Is(z)\times\m\right), 
\end{equation}
where $\m(z,t)$ is the magnetization unit vector; $\gamma(>0)$ is the gyromagnetic ratio; $M_s$ is the saturation magnetization of the magnet; $\mu_0$ is the permeability of free space; $\th$ is the thickness of the sample (Fig.~1(b)); $\alpha$ is the Gilbert dissipation constant; $\Is(z)$ is the vertical spin current density loss which is assumed to be dependent only on $z$; $\Hext$ is the externally applied field; ${\displaystyle \vct{H}_{\text{eff}} = \frac{2\Aex}{\mu_0 M_s} \partial_z^2\m - \Hper m_y + \Hpar m_z}$; $\Aex$ is the exchange energy constant; $\Hper$ is the hard axis anisotropy; $\Hpar$ is the easy axis anisotropy. The above equation can be derived from the Lagrangian along with the generalized forces (see supplementary information) given by the following expressions,
\begin{align}
	L[\theta(z),\phi(z)] &= \left(\mu_0 M_s w \th\right) \int dz \left( -\frac{m_z}{\gamma}\dot{\phi} - \frac{\Aex}{\mu_0 M_s} (\partial_z\m)^2 - \frac\Hper2 m_y^2 + \frac\Hpar2m_z^2 + \m.\Hext \right) \label{Def:Lag} \\
	\delta W &= \left(\mu_0 M_s w \th\right) \int dz \left( -\frac{\alpha}{\gamma} \dot{\m} + \frac{\hbar}{e \mu_0 M_s \th} \left(\Is\times\m\right) \right).\delta\m \label{Def:Force}
\end{align}
where $w$ is the width of the magnet (see Fig.~1(b)). We consider the rigid domain wall ansatz \cite{DW_JOMM}, $\phi(z,t)=\psi(t)$ and ${\displaystyle \theta(z,t) = 2\tan^{-1}\exp\left(\frac{z-Z(t)}{\lambda(\psi)}\right) }$, where the ``width'' of the domain wall, $\lambda(\psi)$, is given by ${\displaystyle \sqrt{\frac{2\Aex}{\mu_0 M_s \left(\Hpar + \Hper\sin^2\psi\right)}} }$. We consider the case of a spin current polarized only along the `$z$' axis with its expression being $\Is = -J^s_z \theta(-z)\uvv{z}$ where $\theta$ is the Heaviside function. Additionally we have a uniform magnetic field along $-z$ direction, $\Hext = -H\uvv{z}$. Finally, we get the equation of motion as,
\begin{align}
	\frac{1}{\lambda} \frac{d Z}{d\tau} &= \Gp \sin\psi\cos\psi + \alpha \frac{d\psi}{d\tau} + \frac\GI2 \frac{1}{1+\exp\left(\frac{2Z}{\lambda}\right)} \label{Eq:RDW:Pos} \\
	\frac{d\psi}{d\tau} &= -\frac12 - \alpha \frac{1}{\lambda}\frac{d Z}{d\tau}, \label{Eq:RDW:Rot}
\end{align}
where the dimensionless time is defined as $\tau = 2\gamma H t$ and we define constants ${\displaystyle \Gp = \frac{\Hper}{2H} }$ and ${\displaystyle \GI = \frac{\hbar J^s_z}{2e\mu_0 M_s H \th} }$. The frequency can directly be deduced from eqs, \eqref{Eq:RDW:Pos} and \eqref{Eq:RDW:Rot}, as follows. Let there exist an oscillatory solution of $Z$ and $\lambda$ with a common dimensionless time ($\tau$) period, say $\Tau$. Note that under such conditions, all the terms in Eq~\eqref{Eq:RDW:Pos} are periodic with period $\Tau$ except possibly the term containing $\sin\psi\cos\psi$. For this term to be periodic with $\Tau$, $\psi$ has to change by an integer multiple of $\pi$ after the time period. Considering Eq~\eqref{Eq:RDW:Rot}, it can be seen that $\psi$ has an oscillatory term along with a drift of ``speed'' $-1/2$. Using the known value of ``speed'' of $\psi$, we can conclude that any oscillations have to exist with a dimensionless time period of $2n\pi$ where $n$ is any natural number. We restrict our attention to $\Tau=2\pi$ which is what we observe in simulation. Coming back to real time, $t$, we conclude that rigid domain wall approximation restricts the angular frequency of oscillation to be $2\gamma H$ (see Eq~\eqref{Eq:Results:Freq}). Note that with this time period, we indeed have an oscillation in $\lambda$ with the same dimensionless time period.

If $\GI=0$, the equation for $\psi$ can be solved exactly\cite{MagDW} and $Z$ can be integrated using Eq~\eqref{Eq:RDW:Rot}. With $\GI\ne0$, there does not seem to be an analytic solution. However, using the intuition that spin current is a small perturbation which mainly acts to negate the effect of dissipation, we can use the solution for field driven motion\cite{MagDW} to approximate the oscillations as,
\begin{equation}\label{Eq:Wave}
	Z(\tau) = Z_C(\GI,\Gp) + \frac{\langle\lambda\rangle}{\alpha} \left[ \tan^{-1}\left(\alpha\Gp + \tan\frac\tau2\right) + 2\pi \text{Ceil}\left(\frac{\tau-\pi}{2\pi}\right) - \frac\tau2 \right]
\end{equation}
where $\text{Ceil}$ is the ceiling function; $Z_C$ is a constant dependent on spin current and fields while the oscillatory part is dependent only on $\Gp$; $\langle \lambda \rangle$ is the average value of $\lambda$ which can be approximated as $\lambda(\psi = \pi/4)$. We can now use Eq~\eqref{Eq:Wave} to analyze the waveform explicitly to calculate various useful observables. First, we use it to find the amplitude of oscillation as ${\displaystyle \frac{2\langle\lambda\rangle}{\alpha}\tan^{-1}\frac{\alpha\Gp}{2} }$. We also note that for $\alpha\Gp\ll1$, this expression reduces approximately to Eq~\eqref{Eq:Results:Amp}. The fact that hard axis anisotropy is responsible for converting rotation into motion is backed up by the expression of the amplitude of the oscillation being approximately proportional to the ratio of the hard axis anisotropy and the magnetic field (see eq~\ref{Eq:Results:Amp}). To calculate the threshold of spin current, we take the time average of Eq~\eqref{Eq:RDW:Pos}. Then using the waveform (see Eq~\eqref{Eq:Wave}) in Eq~\eqref{Eq:RDW:Rot}, we get the inequality, ${\displaystyle \GI>\alpha\left(1+\frac{\Gp^2}{2}\right) }$. This is in accordance with the intuition that the spin current mainly acts to negate the effect of dissipation present in the field driven motion of a domain wall. The results after dismantling the notation are summarised in Eq~\eqref{Eq:Results:Freq}, Eq~\eqref{Eq:Results:Amp} and Eq~\eqref{Eq:Results:Thres}.

\begin{align}
	\omega &= 2\gamma H \label{Eq:Results:Freq}\\
	Z_{\text{max}}-Z_{\text{min}} &\approx \lambda_{eq} \sqrt{\frac{\Hpar}{\Hpar + \Hper/2}} \frac{\Hper}{2H} \label{Eq:Results:Amp}\\
	J^s_{\text{thres}} &\approx \frac{2\alpha e\mu_0 M_s \th}{\hbar} \left( H + \frac{\Hper^2}{8 H} \right). \label{Eq:Results:Thres}
\end{align}

\begin{figure*}
	\includegraphics[width=\textwidth,keepaspectratio]{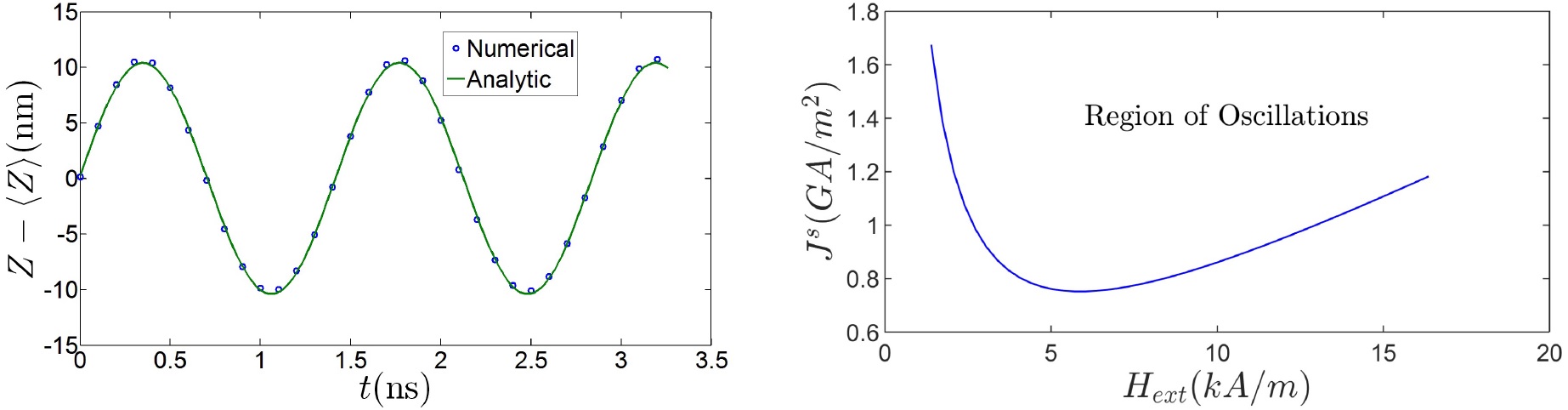}
\caption{Oscillatory waveform and region of occurence (a) Comparison of the oscillatory part of the numerical and the analytic solution of domain wall position vs time, under an external applied field of $10kA/m$. The spin current density for the numerical solution was taken to be $0.96 GA/m^2$. (b) The values of spin current and magnetic field depicting the region in which the oscillations happen. }
\end{figure*}

where $\lambda_{eq}$ is the equilibrium value of $\lambda$ and the factor in the square root (Eq~\eqref{Eq:Results:Amp}) arises due to the variation in the width of the domain wall under oscillatory motion. It will be absent if the variation in width is neglected or in other words, $\Hper\ll\Hpar$. 

We now demonstrate the simulated results of the domain wall motion using the rigid wall approximation discussed above. The waveform derived in Eq.~\eqref{Eq:Wave} is not an exact solution of the equations for the rigid domain wall, but matches fairly well with the numerics as shown in Fig.~2(a). The regime of operation of the device as shown in Fig.~2(b) demonstrates that we need a minimum magnitude of the spin current to compete against the magnetic field and hence result in the oscillations. This can be understood by analyzing the motion of the domain wall when it starts from deep inside either region, i.e., the region deep inside the region of zero or non-zero spin current.
\begin{figure*}
	\includegraphics[width=\textwidth,keepaspectratio]{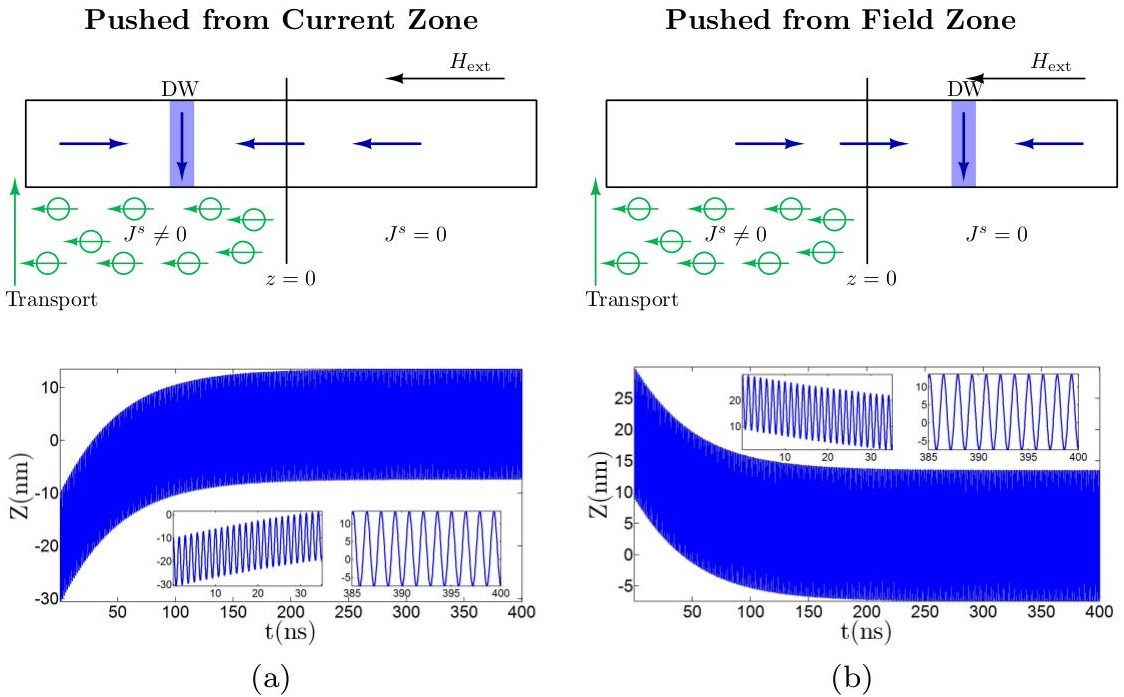}
	\caption{Current push and Field push. (a) Simulation results for domain wall position, $Z$ when started from a point inside the region of non-zero spin current. (b) Similar plot as (a) except for the initial position of domain wall being inside the region of zero spin current.}
\end{figure*}

In Fig.~3(a), we demonstrate the simulated motion of a rigid domain wall starting from a point $z<0$. As shown in the figure, the domain wall will be pushed away until the ``force'' of the spin current is small enough to be compensated by the drift caused by the magnetic field. An opposite scenario is shown in Fig.~3(b), where the domain wall starting deep inside the region of zero spin current ($z>0$) will have a field driven drift until it encounters the region of non-zero spin current. In both the scenarios, the spin current magnitude should be large enough to push back the domain wall or the latter will continue to move indefinitely against the field. 

To verify the results, we have also performed micromagnetic simulations, details of which are included in the supplementary material. We consider a $3\text{nm}$ thick magnetic film with a cross-section of $800\times100\text{nm}$. We assume the magnet parameters, $M_s=8\times10^5\ \text{A}/\text{m}$ and $\Aex=13 pJ/m$; where $M_s$ is the saturation magnetization and $\Aex$ is the exchange energy. We assume a crystalline anisotropy, with its contribution to energy density as $-K_am_y^2$ where $K_a=0.35 \text{MJ}/\text{m}^3$, in the direction of its thickness which works toward reducing the hard axis anisotropy caused by the dipolar interaction. We apply a magnetic field of $8.75\ \text{kA}/\text{m}$ and a spin current density of $20\ \text{GA}/\text{m}^2$. We have simulated it for $40 \text{ns}$ to find that oscillations occur at a frequency close to $0.56 \text{GHz}$. This value is close to the one derived using the theoretical analysis as written in Eq~\eqref{Eq:Results:Freq}.

{\it{Noise Analysis:}} Finally, we verify the stability of the oscillations by adding noise to the rigid domain wall equations and numerically calculating its quality factor. For simplicity, we assume $\Hper \ll \Hpar$ which corresponds to a domain wall with constant width. We introduce two uncorrelated white noise sources, $N_Z$ and $N_\psi$, both of which satisfy $\left\langle N(t) \right\rangle = 0$ and $\left\langle N(t) N(t') \right\rangle = 2\alpha k_B T \delta(t-t')$. The noise can then be added as\cite{NoiseDuine},
\begin{align}
	\frac{1}{\lambda} \frac{d Z}{d\tau} &= \Gp \sin\psi\cos\psi + \alpha \frac{d\psi}{d\tau} + \frac\GI2 \frac{1}{1+\exp\left(\frac{2Z}{\lambda}\right)} \label{Eq:NoisyRDW:Pos} + \frac{N_Z}{\sqrt{\GN}} \\
	\frac{d\psi}{d\tau} &= -\frac12 - \alpha \frac{1}{\lambda}\frac{d Z}{d\tau} + \frac{N_\psi}{\sqrt{\GN}} \label{Eq:NoisyRDW:Rot}
\end{align}
where ${\displaystyle \GN = 4 \gamma \mu_0 M_s H^2 w\th \lambda }$ has been chosen such that Fokker-Planck equation corresponding to Langevin equations Eq~\eqref{Eq:NoisyRDW:Pos} and Eq~\eqref{Eq:NoisyRDW:Rot} admits the Boltzmann distribution in steady state. We simulated the above equations at a room temperature of $300K$ for $40\mu s$ for various values of spin current and magnetic field. The power spectral density (PSD) of $Z$ for three values of the applied magnetic field is plotted in Fig.~4. From the simulated spectrum we find that the quality factor of the oscillator is $\sim550$, $\sim1100$ and $\sim1400$ respectively for the applied fields of $8 kA/m$, $10 kA/m$ and $12 kA/m$ respectively.
\begin{figure*}
	\includegraphics[width=.5\textwidth,keepaspectratio]{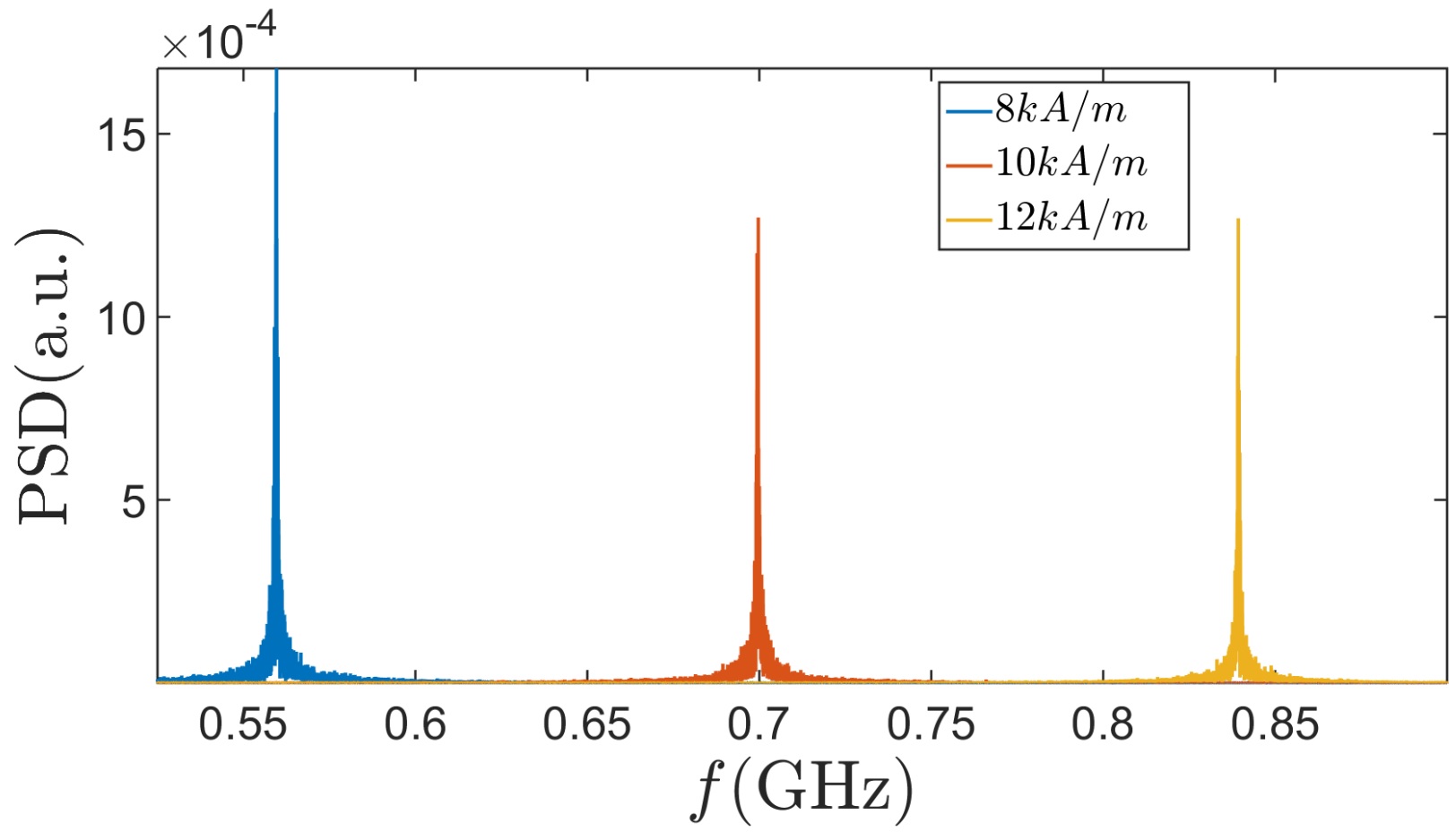}
	\caption{Noise Analysis. Power Spectral Density of the domain wall position for three values of the applied field and with the spin current density fixed at $0.96 GA/m^2$. It shows peaks at the expected values of the oscillation frequency with small width due to noise.}
\end{figure*}

In conclusion, we have proposed a new set up for an oscillator based on the self oscillations of a magnetic domain wall. We found that under rigid domain wall approximation, the oscillatory part of waveform is almost independent of input spin current and the frequency of oscillations is solely governed by the external magnetic field. We also demonstrated a high quality factor giving evidence for the stability of the oscillations. We envision that the simple set up proposed, namely a domain wall subject to a non-uniform vertical spin current will also open up many possibilities for simultaneous write and read out along with the possibility of an all domain wall logic scheme.

{\it{Acknowledgements:}} This work was supported in part by the IIT Bombay SEED grant and the Department of Science and Technology (DST), India, under the Science and Engineering Board grant no. SERB/F/3370/2013-2014. We would also like to acknowledge the support of the center of excellence in nanoelectronics, IIT Bombay. \\
{\it{Competing Interests:}} The authors certify that there are no competing financial interests.\\
{\it{Author Contributions:}}  A.T. conceived the idea. S.S., B. M. and A. T. contributed to the mathematical analysis, simulations, development of the blue prints and the writing of the paper. S.S., B.M. and A.T. extensively discussed the science, the simulations and the results obtained.

\bibliography{References}

\begin{thebibliography}{33}%
\makeatletter
\providecommand \@ifxundefined [1]{%
 \@ifx{#1\undefined}
}%
\providecommand \@ifnum [1]{%
 \ifnum #1\expandafter \@firstoftwo
 \else \expandafter \@secondoftwo
 \fi
}%
\providecommand \@ifx [1]{%
 \ifx #1\expandafter \@firstoftwo
 \else \expandafter \@secondoftwo
 \fi
}%
\providecommand \natexlab [1]{#1}%
\providecommand \enquote  [1]{``#1''}%
\providecommand \bibnamefont  [1]{#1}%
\providecommand \bibfnamefont [1]{#1}%
\providecommand \citenamefont [1]{#1}%
\providecommand \href@noop [0]{\@secondoftwo}%
\providecommand \href [0]{\begingroup \@sanitize@url \@href}%
\providecommand \@href[1]{\@@startlink{#1}\@@href}%
\providecommand \@@href[1]{\endgroup#1\@@endlink}%
\providecommand \@sanitize@url [0]{\catcode `\\12\catcode `\$12\catcode
  `\&12\catcode `\#12\catcode `\^12\catcode `\_12\catcode `\%12\relax}%
\providecommand \@@startlink[1]{}%
\providecommand \@@endlink[0]{}%
\providecommand \url  [0]{\begingroup\@sanitize@url \@url }%
\providecommand \@url [1]{\endgroup\@href {#1}{\urlprefix }}%
\providecommand \urlprefix  [0]{URL }%
\providecommand \Eprint [0]{\href }%
\providecommand \doibase [0]{http://dx.doi.org/}%
\providecommand \selectlanguage [0]{\@gobble}%
\providecommand \bibinfo  [0]{\@secondoftwo}%
\providecommand \bibfield  [0]{\@secondoftwo}%
\providecommand \translation [1]{[#1]}%
\providecommand \BibitemOpen [0]{}%
\providecommand \bibitemStop [0]{}%
\providecommand \bibitemNoStop [0]{.\EOS\space}%
\providecommand \EOS [0]{\spacefactor3000\relax}%
\providecommand \BibitemShut  [1]{\csname bibitem#1\endcsname}%
\let\auto@bib@innerbib\@empty
\bibitem [{\citenamefont {McIntyre}\ \emph {et~al.}(1983)\citenamefont
  {McIntyre}, \citenamefont {Schumacher},\ and\ \citenamefont
  {Woodhouse}}]{Music}%
  \BibitemOpen
  \bibfield  {author} {\bibinfo {author} {\bibfnamefont {M.~E.}\ \bibnamefont
  {McIntyre}}, \bibinfo {author} {\bibfnamefont {R.~T.}\ \bibnamefont
  {Schumacher}}, \ and\ \bibinfo {author} {\bibfnamefont {J.}~\bibnamefont
  {Woodhouse}},\ }\href@noop {} {\bibfield  {journal} {\bibinfo  {journal} {The
  Journal of the Acoustical Society of America}\ }\textbf {\bibinfo {volume}
  {74}},\ \bibinfo {pages} {1325} (\bibinfo {year} {1983})}\BibitemShut
  {NoStop}%
\bibitem [{\citenamefont {Billah}\ and\ \citenamefont
  {Scanlan}(1991)}]{Tacoma}%
  \BibitemOpen
  \bibfield  {author} {\bibinfo {author} {\bibfnamefont {K.~Y.}\ \bibnamefont
  {Billah}}\ and\ \bibinfo {author} {\bibfnamefont {R.~H.}\ \bibnamefont
  {Scanlan}},\ }\href@noop {} {\bibfield  {journal} {\bibinfo  {journal}
  {American Journal of Physics}\ }\textbf {\bibinfo {volume} {59}},\ \bibinfo
  {pages} {118} (\bibinfo {year} {1991})}\BibitemShut {NoStop}%
\bibitem [{\citenamefont {Kiselev}\ \emph {et~al.}(2003)\citenamefont
  {Kiselev}, \citenamefont {Sankey}, \citenamefont {Krivorotov}, \citenamefont
  {Emley}, \citenamefont {Schoelkopf}, \citenamefont {Buhrman},\ and\
  \citenamefont {Ralph}}]{kise}%
  \BibitemOpen
  \bibfield  {author} {\bibinfo {author} {\bibfnamefont {S.~I.}\ \bibnamefont
  {Kiselev}}, \bibinfo {author} {\bibfnamefont {J.}~\bibnamefont {Sankey}},
  \bibinfo {author} {\bibfnamefont {I.}~\bibnamefont {Krivorotov}}, \bibinfo
  {author} {\bibfnamefont {N.}~\bibnamefont {Emley}}, \bibinfo {author}
  {\bibfnamefont {R.}~\bibnamefont {Schoelkopf}}, \bibinfo {author}
  {\bibfnamefont {R.}~\bibnamefont {Buhrman}}, \ and\ \bibinfo {author}
  {\bibfnamefont {D.}~\bibnamefont {Ralph}},\ }\href@noop {} {\bibfield
  {journal} {\bibinfo  {journal} {Nature}\ }\textbf {\bibinfo {volume} {425}},\
  \bibinfo {pages} {380} (\bibinfo {year} {2003})}\BibitemShut {NoStop}%
\bibitem [{\citenamefont {Kim}(2012)}]{Kim_STO}%
  \BibitemOpen
  \bibfield  {author} {\bibinfo {author} {\bibfnamefont {J.-V.}\ \bibnamefont
  {Kim}}\ }(\bibinfo  {publisher} {Academic Press},\ \bibinfo {year} {2012})\
  pp.\ \bibinfo {pages} {217 -- 294}\BibitemShut {NoStop}%
\bibitem [{\citenamefont {Locatelli}\ \emph {et~al.}(2014)\citenamefont
  {Locatelli}, \citenamefont {Cros},\ and\ \citenamefont
  {Grollier}}]{Grollier}%
  \BibitemOpen
  \bibfield  {author} {\bibinfo {author} {\bibfnamefont {N.}~\bibnamefont
  {Locatelli}}, \bibinfo {author} {\bibfnamefont {V.}~\bibnamefont {Cros}}, \
  and\ \bibinfo {author} {\bibfnamefont {J.}~\bibnamefont {Grollier}},\ }\href
  {\doibase DOI:10.1038/nmat3823} {\bibfield  {journal} {\bibinfo  {journal}
  {Nat Mater}\ }\textbf {\bibinfo {volume} {13}},\ \bibinfo {pages} {11}
  (\bibinfo {year} {2014})}\BibitemShut {NoStop}%
\bibitem [{\citenamefont {Ono}\ and\ \citenamefont {Tarucha}(2004)}]{Ono}%
  \BibitemOpen
  \bibfield  {author} {\bibinfo {author} {\bibfnamefont {K.}~\bibnamefont
  {Ono}}\ and\ \bibinfo {author} {\bibfnamefont {S.}~\bibnamefont {Tarucha}},\
  }\href@noop {} {\bibfield  {journal} {\bibinfo  {journal} {Phys. Rev. Lett.}\
  }\textbf {\bibinfo {volume} {92}},\ \bibinfo {pages} {256803} (\bibinfo
  {year} {2004})}\BibitemShut {NoStop}%
\bibitem [{\citenamefont {Takagi}\ and\ \citenamefont {Tatara}(1996)}]{QCoh}%
  \BibitemOpen
  \bibfield  {author} {\bibinfo {author} {\bibfnamefont {S.}~\bibnamefont
  {Takagi}}\ and\ \bibinfo {author} {\bibfnamefont {G.}~\bibnamefont
  {Tatara}},\ }\href@noop {} {\bibfield  {journal} {\bibinfo  {journal} {Phys.
  Rev. B}\ }\textbf {\bibinfo {volume} {54}},\ \bibinfo {pages} {9920}
  (\bibinfo {year} {1996})}\BibitemShut {NoStop}%
\bibitem [{\citenamefont {Braun}\ and\ \citenamefont {Loss}(1996)}]{Soliton}%
  \BibitemOpen
  \bibfield  {author} {\bibinfo {author} {\bibfnamefont {H.-B.}\ \bibnamefont
  {Braun}}\ and\ \bibinfo {author} {\bibfnamefont {D.}~\bibnamefont {Loss}},\
  }\href@noop {} {\bibfield  {journal} {\bibinfo  {journal} {Phys. Rev. B}\
  }\textbf {\bibinfo {volume} {53}},\ \bibinfo {pages} {3237} (\bibinfo {year}
  {1996})}\BibitemShut {NoStop}%
\bibitem [{\citenamefont {Schryer}\ and\ \citenamefont {Walker}(1974)}]{MagDW}%
  \BibitemOpen
  \bibfield  {author} {\bibinfo {author} {\bibfnamefont {N.~L.}\ \bibnamefont
  {Schryer}}\ and\ \bibinfo {author} {\bibfnamefont {L.~R.}\ \bibnamefont
  {Walker}},\ }\href@noop {} {\bibfield  {journal} {\bibinfo  {journal}
  {Journal of Applied Physics}\ }\textbf {\bibinfo {volume} {45}},\ \bibinfo
  {pages} {5406} (\bibinfo {year} {1974})}\BibitemShut {NoStop}%
\bibitem [{\citenamefont {Bouzidi}\ and\ \citenamefont {Suhl}(1990)}]{MagDW2}%
  \BibitemOpen
  \bibfield  {author} {\bibinfo {author} {\bibfnamefont {D.}~\bibnamefont
  {Bouzidi}}\ and\ \bibinfo {author} {\bibfnamefont {H.}~\bibnamefont {Suhl}},\
  }\href@noop {} {\bibfield  {journal} {\bibinfo  {journal} {Phys. Rev. Lett.}\
  }\textbf {\bibinfo {volume} {65}},\ \bibinfo {pages} {2587} (\bibinfo {year}
  {1990})}\BibitemShut {NoStop}%
\bibitem [{\citenamefont {Lucassen}\ \emph {et~al.}(2009)\citenamefont
  {Lucassen}, \citenamefont {van Driel}, \citenamefont {Smith},\ and\
  \citenamefont {Duine}}]{MagNoise}%
  \BibitemOpen
  \bibfield  {author} {\bibinfo {author} {\bibfnamefont {M.~E.}\ \bibnamefont
  {Lucassen}}, \bibinfo {author} {\bibfnamefont {H.~J.}\ \bibnamefont {van
  Driel}}, \bibinfo {author} {\bibfnamefont {C.~M.}\ \bibnamefont {Smith}}, \
  and\ \bibinfo {author} {\bibfnamefont {R.~A.}\ \bibnamefont {Duine}},\
  }\href@noop {} {\bibfield  {journal} {\bibinfo  {journal} {Phys. Rev. B}\
  }\textbf {\bibinfo {volume} {79}},\ \bibinfo {pages} {224411} (\bibinfo
  {year} {2009})}\BibitemShut {NoStop}%
\bibitem [{\citenamefont {Gorchon}\ \emph {et~al.}(2014)\citenamefont
  {Gorchon}, \citenamefont {Bustingorry}, \citenamefont {Ferr'e}, \citenamefont
  {Jeudy}, \citenamefont {Kolton},\ and\ \citenamefont {Giamarchi}}]{PinMagDW}%
  \BibitemOpen
  \bibfield  {author} {\bibinfo {author} {\bibfnamefont {J.}~\bibnamefont
  {Gorchon}}, \bibinfo {author} {\bibfnamefont {S.}~\bibnamefont
  {Bustingorry}}, \bibinfo {author} {\bibfnamefont {J.}~\bibnamefont {Ferr'e}},
  \bibinfo {author} {\bibfnamefont {V.}~\bibnamefont {Jeudy}}, \bibinfo
  {author} {\bibfnamefont {B.}~\bibnamefont {Kolton}, \bibfnamefont {A.}}, \
  and\ \bibinfo {author} {\bibfnamefont {T.}~\bibnamefont {Giamarchi}},\
  }\href@noop {} {\bibfield  {journal} {\bibinfo  {journal} {Phys. Rev. Lett.}\
  }\textbf {\bibinfo {volume} {113}},\ \bibinfo {pages} {027205} (\bibinfo
  {year} {2014})}\BibitemShut {NoStop}%
\bibitem [{\citenamefont {Saitoh}\ \emph {et~al.}(2004)\citenamefont {Saitoh},
  \citenamefont {Miyajima}, \citenamefont {Yamaoka},\ and\ \citenamefont
  {Tatara}}]{CurDWExp}%
  \BibitemOpen
  \bibfield  {author} {\bibinfo {author} {\bibfnamefont {E.}~\bibnamefont
  {Saitoh}}, \bibinfo {author} {\bibfnamefont {H.}~\bibnamefont {Miyajima}},
  \bibinfo {author} {\bibfnamefont {T.}~\bibnamefont {Yamaoka}}, \ and\
  \bibinfo {author} {\bibfnamefont {G.}~\bibnamefont {Tatara}},\ }\href@noop {}
  {\bibfield  {journal} {\bibinfo  {journal} {Nature}\ }\textbf {\bibinfo
  {volume} {432}},\ \bibinfo {pages} {203} (\bibinfo {year}
  {2004})}\BibitemShut {NoStop}%
\bibitem [{\citenamefont {Yamaguchi}\ \emph {et~al.}(2004)\citenamefont
  {Yamaguchi}, \citenamefont {Ono}, \citenamefont {Nasu}, \citenamefont
  {Miyake}, \citenamefont {Mibu},\ and\ \citenamefont {Shinjo}}]{CurDWExp2}%
  \BibitemOpen
  \bibfield  {author} {\bibinfo {author} {\bibfnamefont {A.}~\bibnamefont
  {Yamaguchi}}, \bibinfo {author} {\bibfnamefont {T.}~\bibnamefont {Ono}},
  \bibinfo {author} {\bibfnamefont {S.}~\bibnamefont {Nasu}}, \bibinfo {author}
  {\bibfnamefont {K.}~\bibnamefont {Miyake}}, \bibinfo {author} {\bibfnamefont
  {K.}~\bibnamefont {Mibu}}, \ and\ \bibinfo {author} {\bibfnamefont
  {T.}~\bibnamefont {Shinjo}},\ }\href@noop {} {\bibfield  {journal} {\bibinfo
  {journal} {Phys. Rev. Lett.}\ }\textbf {\bibinfo {volume} {92}},\ \bibinfo
  {pages} {077205} (\bibinfo {year} {2004})}\BibitemShut {NoStop}%
\bibitem [{\citenamefont {Tatara}\ and\ \citenamefont {Kohno}(2004)}]{CurDWTh}%
  \BibitemOpen
  \bibfield  {author} {\bibinfo {author} {\bibfnamefont {G.}~\bibnamefont
  {Tatara}}\ and\ \bibinfo {author} {\bibfnamefont {H.}~\bibnamefont {Kohno}},\
  }\href@noop {} {\bibfield  {journal} {\bibinfo  {journal} {Phys. Rev. Lett.}\
  }\textbf {\bibinfo {volume} {92}},\ \bibinfo {pages} {086601} (\bibinfo
  {year} {2004})}\BibitemShut {NoStop}%
\bibitem [{\citenamefont {Catalan}\ \emph {et~al.}(2012)\citenamefont
  {Catalan}, \citenamefont {Seidel}, \citenamefont {Ramesh},\ and\
  \citenamefont {Scott}}]{Rev_Mod_DW}%
  \BibitemOpen
  \bibfield  {author} {\bibinfo {author} {\bibfnamefont {G.}~\bibnamefont
  {Catalan}}, \bibinfo {author} {\bibfnamefont {J.}~\bibnamefont {Seidel}},
  \bibinfo {author} {\bibfnamefont {R.}~\bibnamefont {Ramesh}}, \ and\ \bibinfo
  {author} {\bibfnamefont {J.~F.}\ \bibnamefont {Scott}},\ }\href@noop {}
  {\bibfield  {journal} {\bibinfo  {journal} {Rev. Mod. Phys.}\ }\textbf
  {\bibinfo {volume} {84}},\ \bibinfo {pages} {119} (\bibinfo {year}
  {2012})}\BibitemShut {NoStop}%
\bibitem [{\citenamefont {Parkin}\ \emph {et~al.}(2008)\citenamefont {Parkin},
  \citenamefont {Hayashi},\ and\ \citenamefont {Thomas}}]{RaceTrack}%
  \BibitemOpen
  \bibfield  {author} {\bibinfo {author} {\bibfnamefont {S.~S.~P.}\
  \bibnamefont {Parkin}}, \bibinfo {author} {\bibfnamefont {M.}~\bibnamefont
  {Hayashi}}, \ and\ \bibinfo {author} {\bibfnamefont {L.}~\bibnamefont
  {Thomas}},\ }\href@noop {} {\bibfield  {journal} {\bibinfo  {journal}
  {Science}\ }\textbf {\bibinfo {volume} {320}},\ \bibinfo {pages} {190}
  (\bibinfo {year} {2008})}\BibitemShut {NoStop}%
\bibitem [{\citenamefont {Yang}\ \emph {et~al.}(2015)\citenamefont {Yang},
  \citenamefont {Ryu},\ and\ \citenamefont {Parkin}}]{ExpDW}%
  \BibitemOpen
  \bibfield  {author} {\bibinfo {author} {\bibfnamefont {S.-H.}\ \bibnamefont
  {Yang}}, \bibinfo {author} {\bibfnamefont {K.-S.}\ \bibnamefont {Ryu}}, \
  and\ \bibinfo {author} {\bibfnamefont {S.}~\bibnamefont {Parkin}},\
  }\href@noop {} {\bibfield  {journal} {\bibinfo  {journal} {Nat Nano}\
  }\textbf {\bibinfo {volume} {10}},\ \bibinfo {pages} {221} (\bibinfo {year}
  {2015})}\BibitemShut {NoStop}%
\bibitem [{\citenamefont {Yang}\ \emph {et~al.}(2008)\citenamefont {Yang},
  \citenamefont {Nistor}, \citenamefont {Beach},\ and\ \citenamefont
  {Erskine}}]{MagOsc}%
  \BibitemOpen
  \bibfield  {author} {\bibinfo {author} {\bibfnamefont {J.}~\bibnamefont
  {Yang}}, \bibinfo {author} {\bibfnamefont {C.}~\bibnamefont {Nistor}},
  \bibinfo {author} {\bibfnamefont {G.~S.~D.}\ \bibnamefont {Beach}}, \ and\
  \bibinfo {author} {\bibfnamefont {J.~L.}\ \bibnamefont {Erskine}},\
  }\href@noop {} {\bibfield  {journal} {\bibinfo  {journal} {Phys. Rev. B}\
  }\textbf {\bibinfo {volume} {77}},\ \bibinfo {pages} {014413} (\bibinfo
  {year} {2008})}\BibitemShut {NoStop}%
\bibitem [{\citenamefont {Chanthbouala}\ \emph {et~al.}(2011)\citenamefont
  {Chanthbouala}, \citenamefont {Matsumoto}, \citenamefont {Grollier},
  \citenamefont {Cros}, \citenamefont {Anane}, \citenamefont {Fert},
  \citenamefont {Khvalkovskiy}, \citenamefont {Zvezdin}, \citenamefont
  {Nishimura}, \citenamefont {Nagamine}, \citenamefont {Maehara}, \citenamefont
  {Tsunekawa}, \citenamefont {Fukushima},\ and\ \citenamefont
  {Yuasa}}]{VerCurExp1}%
  \BibitemOpen
  \bibfield  {author} {\bibinfo {author} {\bibfnamefont {A.}~\bibnamefont
  {Chanthbouala}}, \bibinfo {author} {\bibfnamefont {R.}~\bibnamefont
  {Matsumoto}}, \bibinfo {author} {\bibfnamefont {J.}~\bibnamefont {Grollier}},
  \bibinfo {author} {\bibfnamefont {V.}~\bibnamefont {Cros}}, \bibinfo {author}
  {\bibfnamefont {A.}~\bibnamefont {Anane}}, \bibinfo {author} {\bibfnamefont
  {A.}~\bibnamefont {Fert}}, \bibinfo {author} {\bibfnamefont {A.~V.}\
  \bibnamefont {Khvalkovskiy}}, \bibinfo {author} {\bibfnamefont {K.~A.}\
  \bibnamefont {Zvezdin}}, \bibinfo {author} {\bibfnamefont {K.}~\bibnamefont
  {Nishimura}}, \bibinfo {author} {\bibfnamefont {Y.}~\bibnamefont {Nagamine}},
  \bibinfo {author} {\bibfnamefont {H.}~\bibnamefont {Maehara}}, \bibinfo
  {author} {\bibfnamefont {K.}~\bibnamefont {Tsunekawa}}, \bibinfo {author}
  {\bibfnamefont {A.}~\bibnamefont {Fukushima}}, \ and\ \bibinfo {author}
  {\bibfnamefont {S.}~\bibnamefont {Yuasa}},\ }\href@noop {} {\bibfield
  {journal} {\bibinfo  {journal} {Nat Phys}\ }\textbf {\bibinfo {volume} {7}},\
  \bibinfo {pages} {626} (\bibinfo {year} {2011})}\BibitemShut {NoStop}%
\bibitem [{\citenamefont {Metaxas}\ \emph {et~al.}(2013)\citenamefont
  {Metaxas}, \citenamefont {Sampaio}, \citenamefont {Chanthbouala},
  \citenamefont {Matsumoto}, \citenamefont {Anane}, \citenamefont {Fert},
  \citenamefont {Zvezdin}, \citenamefont {Yakushiji}, \citenamefont {Kubota},
  \citenamefont {Fukushima}, \citenamefont {Yuasa}, \citenamefont {Nishimura},
  \citenamefont {Nagamine}, \citenamefont {Maehara}, \citenamefont {Tsunekawa},
  \citenamefont {Cros},\ and\ \citenamefont {Grollier}}]{VerCurExp}%
  \BibitemOpen
  \bibfield  {author} {\bibinfo {author} {\bibfnamefont {P.~J.}\ \bibnamefont
  {Metaxas}}, \bibinfo {author} {\bibfnamefont {J.}~\bibnamefont {Sampaio}},
  \bibinfo {author} {\bibfnamefont {A.}~\bibnamefont {Chanthbouala}}, \bibinfo
  {author} {\bibfnamefont {R.}~\bibnamefont {Matsumoto}}, \bibinfo {author}
  {\bibfnamefont {A.}~\bibnamefont {Anane}}, \bibinfo {author} {\bibfnamefont
  {A.}~\bibnamefont {Fert}}, \bibinfo {author} {\bibfnamefont {K.~A.}\
  \bibnamefont {Zvezdin}}, \bibinfo {author} {\bibfnamefont {K.}~\bibnamefont
  {Yakushiji}}, \bibinfo {author} {\bibfnamefont {H.}~\bibnamefont {Kubota}},
  \bibinfo {author} {\bibfnamefont {A.}~\bibnamefont {Fukushima}}, \bibinfo
  {author} {\bibfnamefont {S.}~\bibnamefont {Yuasa}}, \bibinfo {author}
  {\bibfnamefont {K.}~\bibnamefont {Nishimura}}, \bibinfo {author}
  {\bibfnamefont {Y.}~\bibnamefont {Nagamine}}, \bibinfo {author}
  {\bibfnamefont {H.}~\bibnamefont {Maehara}}, \bibinfo {author} {\bibfnamefont
  {K.}~\bibnamefont {Tsunekawa}}, \bibinfo {author} {\bibfnamefont
  {V.}~\bibnamefont {Cros}}, \ and\ \bibinfo {author} {\bibfnamefont
  {J.}~\bibnamefont {Grollier}},\ }\href@noop {} {\bibfield  {journal}
  {\bibinfo  {journal} {Sci. Rep.}\ }\textbf {\bibinfo {volume} {3}},\ \bibinfo
  {pages} {DOI:10.1038/srep01829} (\bibinfo {year} {2013})}\BibitemShut
  {NoStop}%
\bibitem [{\citenamefont {Khvalkovskiy}\ \emph {et~al.}(2009)\citenamefont
  {Khvalkovskiy}, \citenamefont {Zvezdin}, \citenamefont {Gorbunov},
  \citenamefont {Cros}, \citenamefont {Grollier}, \citenamefont {Fert},\ and\
  \citenamefont {Zvezdin}}]{VerCurTh}%
  \BibitemOpen
  \bibfield  {author} {\bibinfo {author} {\bibfnamefont {A.~V.}\ \bibnamefont
  {Khvalkovskiy}}, \bibinfo {author} {\bibfnamefont {K.~A.}\ \bibnamefont
  {Zvezdin}}, \bibinfo {author} {\bibfnamefont {Y.~V.}\ \bibnamefont
  {Gorbunov}}, \bibinfo {author} {\bibfnamefont {V.}~\bibnamefont {Cros}},
  \bibinfo {author} {\bibfnamefont {J.}~\bibnamefont {Grollier}}, \bibinfo
  {author} {\bibfnamefont {A.}~\bibnamefont {Fert}}, \ and\ \bibinfo {author}
  {\bibfnamefont {A.~K.}\ \bibnamefont {Zvezdin}},\ }\href@noop {} {\bibfield
  {journal} {\bibinfo  {journal} {Phys. Rev. Lett.}\ }\textbf {\bibinfo
  {volume} {102}},\ \bibinfo {pages} {067206} (\bibinfo {year}
  {2009})}\BibitemShut {NoStop}%
\bibitem [{\citenamefont {Slonczewski}(1996)}]{Slon}%
  \BibitemOpen
  \bibfield  {author} {\bibinfo {author} {\bibfnamefont {J.}~\bibnamefont
  {Slonczewski}},\ }\href@noop {} {\bibfield  {journal} {\bibinfo  {journal}
  {Journal of Magnetism and Magnetic Materials}\ }\textbf {\bibinfo {volume}
  {159}},\ \bibinfo {pages} {L1 } (\bibinfo {year} {1996})}\BibitemShut
  {NoStop}%
\bibitem [{\citenamefont {Zhang}\ \emph {et~al.}(2002)\citenamefont {Zhang},
  \citenamefont {Levy},\ and\ \citenamefont {Fert}}]{MechSTT}%
  \BibitemOpen
  \bibfield  {author} {\bibinfo {author} {\bibfnamefont {S.}~\bibnamefont
  {Zhang}}, \bibinfo {author} {\bibfnamefont {P.~M.}\ \bibnamefont {Levy}}, \
  and\ \bibinfo {author} {\bibfnamefont {A.}~\bibnamefont {Fert}},\ }\href@noop
  {} {\bibfield  {journal} {\bibinfo  {journal} {Phys. Rev. Lett.}\ }\textbf
  {\bibinfo {volume} {88}},\ \bibinfo {pages} {236601} (\bibinfo {year}
  {2002})}\BibitemShut {NoStop}%
\bibitem [{\citenamefont {Tulapurkar}\ \emph {et~al.}(2005)\citenamefont
  {Tulapurkar}, \citenamefont {Suzuki}, \citenamefont {Fukushima},
  \citenamefont {Kubota}, \citenamefont {Maehara}, \citenamefont {Tsunekawa},
  \citenamefont {Djayaprawira}, \citenamefont {Watanabe},\ and\ \citenamefont
  {Yuasa}}]{STDiode}%
  \BibitemOpen
  \bibfield  {author} {\bibinfo {author} {\bibfnamefont {A.~A.}\ \bibnamefont
  {Tulapurkar}}, \bibinfo {author} {\bibfnamefont {Y.}~\bibnamefont {Suzuki}},
  \bibinfo {author} {\bibfnamefont {A.}~\bibnamefont {Fukushima}}, \bibinfo
  {author} {\bibfnamefont {H.}~\bibnamefont {Kubota}}, \bibinfo {author}
  {\bibfnamefont {H.}~\bibnamefont {Maehara}}, \bibinfo {author} {\bibfnamefont
  {K.}~\bibnamefont {Tsunekawa}}, \bibinfo {author} {\bibfnamefont {D.~D.}\
  \bibnamefont {Djayaprawira}}, \bibinfo {author} {\bibfnamefont
  {N.}~\bibnamefont {Watanabe}}, \ and\ \bibinfo {author} {\bibfnamefont
  {S.}~\bibnamefont {Yuasa}},\ }\href@noop {} {\bibfield  {journal} {\bibinfo
  {journal} {Nature}\ }\textbf {\bibinfo {volume} {438}},\ \bibinfo {pages}
  {339} (\bibinfo {year} {2005})}\BibitemShut {NoStop}%
\bibitem [{\citenamefont {Ralph}\ and\ \citenamefont
  {Stiles}(2007)}]{Ralph_STT}%
  \BibitemOpen
  \bibfield  {author} {\bibinfo {author} {\bibfnamefont {D.}~\bibnamefont
  {Ralph}}\ and\ \bibinfo {author} {\bibfnamefont {M.}~\bibnamefont {Stiles}},\
  }\href@noop {} {\bibfield  {journal} {\bibinfo  {journal} {Journal of
  Magnetism and Magnetic Materials}\ }\textbf {\bibinfo {volume} {320}},\
  \bibinfo {pages} {1190} (\bibinfo {year} {2007})}\BibitemShut {NoStop}%
\bibitem [{\citenamefont {Zhang}\ \emph {et~al.}(2013)\citenamefont {Zhang},
  \citenamefont {Wong}, \citenamefont {Yan}, \citenamefont {Wu}, \citenamefont
  {Morton}, \citenamefont {Wang}, \citenamefont {Hu}, \citenamefont {Xu},
  \citenamefont {Scholl}, \citenamefont {Young}, \citenamefont {Barsukov},
  \citenamefont {Farle},\ and\ \citenamefont {van~der Laan}}]{CurOsc_Pos}%
  \BibitemOpen
  \bibfield  {author} {\bibinfo {author} {\bibfnamefont {W.}~\bibnamefont
  {Zhang}}, \bibinfo {author} {\bibfnamefont {P.~K.~J.}\ \bibnamefont {Wong}},
  \bibinfo {author} {\bibfnamefont {P.}~\bibnamefont {Yan}}, \bibinfo {author}
  {\bibfnamefont {J.}~\bibnamefont {Wu}}, \bibinfo {author} {\bibfnamefont
  {S.~A.}\ \bibnamefont {Morton}}, \bibinfo {author} {\bibfnamefont {X.~R.}\
  \bibnamefont {Wang}}, \bibinfo {author} {\bibfnamefont {X.~F.}\ \bibnamefont
  {Hu}}, \bibinfo {author} {\bibfnamefont {Y.~B.}\ \bibnamefont {Xu}}, \bibinfo
  {author} {\bibfnamefont {A.}~\bibnamefont {Scholl}}, \bibinfo {author}
  {\bibfnamefont {A.}~\bibnamefont {Young}}, \bibinfo {author} {\bibfnamefont
  {I.}~\bibnamefont {Barsukov}}, \bibinfo {author} {\bibfnamefont
  {M.}~\bibnamefont {Farle}}, \ and\ \bibinfo {author} {\bibfnamefont
  {G.}~\bibnamefont {van~der Laan}},\ }\href@noop {} {\bibfield  {journal}
  {\bibinfo  {journal} {Applied Physics Letters}\ }\textbf {\bibinfo {volume}
  {103}},\ \bibinfo {eid} {042403} (\bibinfo {year} {2013})}\BibitemShut
  {NoStop}%
\bibitem [{\citenamefont {Martinez}\ \emph {et~al.}(2011)\citenamefont
  {Martinez}, \citenamefont {Torres},\ and\ \citenamefont
  {Lopez-Diaz}}]{StableOsc}%
  \BibitemOpen
  \bibfield  {author} {\bibinfo {author} {\bibfnamefont {E.}~\bibnamefont
  {Martinez}}, \bibinfo {author} {\bibfnamefont {L.}~\bibnamefont {Torres}}, \
  and\ \bibinfo {author} {\bibfnamefont {L.}~\bibnamefont {Lopez-Diaz}},\
  }\href@noop {} {\bibfield  {journal} {\bibinfo  {journal} {Phys. Rev. B}\
  }\textbf {\bibinfo {volume} {83}},\ \bibinfo {pages} {174444} (\bibinfo
  {year} {2011})}\BibitemShut {NoStop}%
\bibitem [{\citenamefont {Khvalkovskiy}\ \emph {et~al.}(2013)\citenamefont
  {Khvalkovskiy}, \citenamefont {Cros}, \citenamefont {Apalkov}, \citenamefont
  {Nikitin}, \citenamefont {Krounbi}, \citenamefont {Zvezdin}, \citenamefont
  {Anane}, \citenamefont {Grollier},\ and\ \citenamefont {Fert}}]{Khav}%
  \BibitemOpen
  \bibfield  {author} {\bibinfo {author} {\bibfnamefont {A.~V.}\ \bibnamefont
  {Khvalkovskiy}}, \bibinfo {author} {\bibfnamefont {V.}~\bibnamefont {Cros}},
  \bibinfo {author} {\bibfnamefont {D.}~\bibnamefont {Apalkov}}, \bibinfo
  {author} {\bibfnamefont {V.}~\bibnamefont {Nikitin}}, \bibinfo {author}
  {\bibfnamefont {M.}~\bibnamefont {Krounbi}}, \bibinfo {author} {\bibfnamefont
  {K.~A.}\ \bibnamefont {Zvezdin}}, \bibinfo {author} {\bibfnamefont
  {A.}~\bibnamefont {Anane}}, \bibinfo {author} {\bibfnamefont
  {J.}~\bibnamefont {Grollier}}, \ and\ \bibinfo {author} {\bibfnamefont
  {A.}~\bibnamefont {Fert}},\ }\href@noop {} {\bibfield  {journal} {\bibinfo
  {journal} {Phys. Rev. B}\ }\textbf {\bibinfo {volume} {87}},\ \bibinfo
  {pages} {DOI:10.1103/PhysRevB.87.020402} (\bibinfo {year}
  {2013})}\BibitemShut {NoStop}%
\bibitem [{\citenamefont {Liu}\ \emph {et~al.}(2012)\citenamefont {Liu},
  \citenamefont {Pai}, \citenamefont {Li}, \citenamefont {Tseng}, \citenamefont
  {Ralph},\ and\ \citenamefont {Buhrman}}]{GSHE_Ta}%
  \BibitemOpen
  \bibfield  {author} {\bibinfo {author} {\bibfnamefont {L.}~\bibnamefont
  {Liu}}, \bibinfo {author} {\bibfnamefont {C.-F.}\ \bibnamefont {Pai}},
  \bibinfo {author} {\bibfnamefont {Y.}~\bibnamefont {Li}}, \bibinfo {author}
  {\bibfnamefont {H.~W.}\ \bibnamefont {Tseng}}, \bibinfo {author}
  {\bibfnamefont {D.~C.}\ \bibnamefont {Ralph}}, \ and\ \bibinfo {author}
  {\bibfnamefont {R.~A.}\ \bibnamefont {Buhrman}},\ }\href@noop {} {\bibfield
  {journal} {\bibinfo  {journal} {Science}\ }\textbf {\bibinfo {volume}
  {336}},\ \bibinfo {pages} {555} (\bibinfo {year} {2012})}\BibitemShut
  {NoStop}%
\bibitem [{\citenamefont {{Bhowmik}}\ \emph {et~al.}(2015)\citenamefont
  {{Bhowmik}}, \citenamefont {{Nowakowski}}, \citenamefont {{You}},
  \citenamefont {{Lee}}, \citenamefont {{Keating}}, \citenamefont {{Wong}},
  \citenamefont {{Bokor}},\ and\ \citenamefont {{Salahuddin}}}]{Ssalahud}%
  \BibitemOpen
  \bibfield  {author} {\bibinfo {author} {\bibfnamefont {D.}~\bibnamefont
  {{Bhowmik}}}, \bibinfo {author} {\bibfnamefont {M.~E.}\ \bibnamefont
  {{Nowakowski}}}, \bibinfo {author} {\bibfnamefont {L.}~\bibnamefont {{You}}},
  \bibinfo {author} {\bibfnamefont {O.}~\bibnamefont {{Lee}}}, \bibinfo
  {author} {\bibfnamefont {D.}~\bibnamefont {{Keating}}}, \bibinfo {author}
  {\bibfnamefont {M.}~\bibnamefont {{Wong}}}, \bibinfo {author} {\bibfnamefont
  {J.}~\bibnamefont {{Bokor}}}, \ and\ \bibinfo {author} {\bibfnamefont
  {S.}~\bibnamefont {{Salahuddin}}},\ }\href@noop {} {\bibfield  {journal}
  {\bibinfo  {journal} {Scientific Reports}\ }\textbf {\bibinfo {volume} {5}},\
  \bibinfo {pages} {10} (\bibinfo {year} {2015})}\BibitemShut {NoStop}%
\bibitem [{\citenamefont {Tserkovnyak}\ \emph {et~al.}(2008)\citenamefont
  {Tserkovnyak}, \citenamefont {Brataas},\ and\ \citenamefont
  {Bauer}}]{DW_JOMM}%
  \BibitemOpen
  \bibfield  {author} {\bibinfo {author} {\bibfnamefont {Y.}~\bibnamefont
  {Tserkovnyak}}, \bibinfo {author} {\bibfnamefont {A.}~\bibnamefont
  {Brataas}}, \ and\ \bibinfo {author} {\bibfnamefont {G.~E.}\ \bibnamefont
  {Bauer}},\ }\href@noop {} {\bibfield  {journal} {\bibinfo  {journal} {Journal
  of Magnetism and Magnetic Materials}\ }\textbf {\bibinfo {volume} {320}},\
  \bibinfo {pages} {1282 } (\bibinfo {year} {2008})}\BibitemShut {NoStop}%
\bibitem [{\citenamefont {Duine}\ \emph {et~al.}(2007)\citenamefont {Duine},
  \citenamefont {N\'u\~nez},\ and\ \citenamefont {MacDonald}}]{NoiseDuine}%
  \BibitemOpen
  \bibfield  {author} {\bibinfo {author} {\bibfnamefont {R.~A.}\ \bibnamefont
  {Duine}}, \bibinfo {author} {\bibfnamefont {A.~S.}\ \bibnamefont
  {N\'u\~nez}}, \ and\ \bibinfo {author} {\bibfnamefont {A.~H.}\ \bibnamefont
  {MacDonald}},\ }\href@noop {} {\bibfield  {journal} {\bibinfo  {journal}
  {Phys. Rev. Lett.}\ }\textbf {\bibinfo {volume} {98}},\ \bibinfo {pages}
  {056605} (\bibinfo {year} {2007})}\BibitemShut {NoStop}%
\end{thebibliography}%


\begin{thebibliography}{4}%
\makeatletter
\providecommand \@ifxundefined [1]{%
 \@ifx{#1\undefined}
}%
\providecommand \@ifnum [1]{%
 \ifnum #1\expandafter \@firstoftwo
 \else \expandafter \@secondoftwo
 \fi
}%
\providecommand \@ifx [1]{%
 \ifx #1\expandafter \@firstoftwo
 \else \expandafter \@secondoftwo
 \fi
}%
\providecommand \natexlab [1]{#1}%
\providecommand \enquote  [1]{``#1''}%
\providecommand \bibnamefont  [1]{#1}%
\providecommand \bibfnamefont [1]{#1}%
\providecommand \citenamefont [1]{#1}%
\providecommand \href@noop [0]{\@secondoftwo}%
\providecommand \href [0]{\begingroup \@sanitize@url \@href}%
\providecommand \@href[1]{\@@startlink{#1}\@@href}%
\providecommand \@@href[1]{\endgroup#1\@@endlink}%
\providecommand \@sanitize@url [0]{\catcode `\\12\catcode `\$12\catcode
  `\&12\catcode `\#12\catcode `\^12\catcode `\_12\catcode `\%12\relax}%
\providecommand \@@startlink[1]{}%
\providecommand \@@endlink[0]{}%
\providecommand \url  [0]{\begingroup\@sanitize@url \@url }%
\providecommand \@url [1]{\endgroup\@href {#1}{\urlprefix }}%
\providecommand \urlprefix  [0]{URL }%
\providecommand \Eprint [0]{\href }%
\providecommand \doibase [0]{http://dx.doi.org/}%
\providecommand \selectlanguage [0]{\@gobble}%
\providecommand \bibinfo  [0]{\@secondoftwo}%
\providecommand \bibfield  [0]{\@secondoftwo}%
\providecommand \translation [1]{[#1]}%
\providecommand \BibitemOpen [0]{}%
\providecommand \bibitemStop [0]{}%
\providecommand \bibitemNoStop [0]{.\EOS\space}%
\providecommand \EOS [0]{\spacefactor3000\relax}%
\providecommand \BibitemShut  [1]{\csname bibitem#1\endcsname}%
\let\auto@bib@innerbib\@empty
\bibitem [{\citenamefont {Slonczewski}(1996)}]{Slon}%
  \BibitemOpen
  \bibfield  {author} {\bibinfo {author} {\bibfnamefont {J.}~\bibnamefont
  {Slonczewski}},\ }\href@noop {} {\bibfield  {journal} {\bibinfo  {journal}
  {Journal of Magnetism and Magnetic Materials}\ }\textbf {\bibinfo {volume}
  {159}},\ \bibinfo {pages} {L1 } (\bibinfo {year} {1996})}\BibitemShut
  {NoStop}%
\bibitem [{\citenamefont {Miltat}\ and\ \citenamefont
  {Donahue}(2007)}]{Miltat}%
  \BibitemOpen
  \bibfield  {author} {\bibinfo {author} {\bibfnamefont {J.~E.}\ \bibnamefont
  {Miltat}}\ and\ \bibinfo {author} {\bibfnamefont {M.~J.}\ \bibnamefont
  {Donahue}},\ }\enquote {\bibinfo {title} {Numerical micromagnetics: Finite
  difference methods},}\ in\ \href {\doibase DOI:10.1002/9780470022184.hmm202}
  {\emph {\bibinfo {booktitle} {Handbook of Magnetism and Advanced Magnetic
  Materials}}}\ (\bibinfo  {publisher} {John Wiley \& Sons, Ltd},\ \bibinfo
  {year} {2007})\BibitemShut {NoStop}%
\bibitem [{\citenamefont {McMichael}\ and\ \citenamefont
  {Donahue}(1997)}]{InfDW}%
  \BibitemOpen
  \bibfield  {author} {\bibinfo {author} {\bibfnamefont {R.}~\bibnamefont
  {McMichael}}\ and\ \bibinfo {author} {\bibfnamefont {M.}~\bibnamefont
  {Donahue}},\ }\href {\doibase DOI:10.1109/20.619698} {\bibfield  {journal}
  {\bibinfo  {journal} {Magnetics, IEEE Transactions on}\ }\textbf {\bibinfo
  {volume} {33}},\ \bibinfo {pages} {4167} (\bibinfo {year}
  {1997})}\BibitemShut {NoStop}%
\bibitem [{\citenamefont {Thiaville}\ \emph {et~al.}(2004)\citenamefont
  {Thiaville}, \citenamefont {Nakatani}, \citenamefont {Miltat},\ and\
  \citenamefont {Vernier}}]{MM_RDW}%
  \BibitemOpen
  \bibfield  {author} {\bibinfo {author} {\bibfnamefont {A.}~\bibnamefont
  {Thiaville}}, \bibinfo {author} {\bibfnamefont {Y.}~\bibnamefont {Nakatani}},
  \bibinfo {author} {\bibfnamefont {J.}~\bibnamefont {Miltat}}, \ and\ \bibinfo
  {author} {\bibfnamefont {N.}~\bibnamefont {Vernier}},\ }\href {\doibase
  DOI:10.1063/1.1667804} {\bibfield  {journal} {\bibinfo  {journal} {Journal of
  Applied Physics}\ }\textbf {\bibinfo {volume} {95}},\ \bibinfo {pages} {7049}
  (\bibinfo {year} {2004})}\BibitemShut {NoStop}%
\end{thebibliography}%
\end{document}


\title{Supplementary Information - Proposal for a Domain Wall Nano-Oscillator driven by Non-uniform Spin Currents}
\author{Sanchar Sharma}
\affiliation{Department of Electrical Engineering, Indian Institute of Technology Bombay, Powai, Mumbai-400076, India}

\author{Bhaskaran Muralidharan*}
\email{bm@ee.iitb.ac.in}
\affiliation{Department of Electrical Engineering, Indian Institute of Technology Bombay, Powai, Mumbai-400076, India}

\author{Ashwin Tulapurkar}
\affiliation{Department of Electrical Engineering, Indian Institute of Technology Bombay, Powai, Mumbai-400076, India}

\begin{abstract}
Here, we provide details about the rigid wall approximation along with the details of the micromagnetic simulations performed in the paper.
\end{abstract}

\flushbottom
\maketitle

\thispagestyle{empty}

\section{Derivation of Lagrangian}

The starting point of magnetization dynamics is the Landau-Lifshitz-Gilbert (LLG) equation augmented with Slonczweski spin torque\cite{Slon} term
\begin{equation} \label{Eq:LLG}
	\frac{d \m}{dt} = -\gamma \m\times \vct{H}_{\text{eff}} + \alpha\m\times\left(\frac{d\m}{dt}\right) - \frac{\gamma \hbar}{e\mu_0 M_s \th}\m\times\left(\Is(z)\times\m\right) 
\end{equation}
where $\Is(z)$ is the incident spin current density which is assumed to be dependent only on $z$; ${\displaystyle \vct{H}_{\text{eff}} = \frac{-1}{\mu_0 M_s} \frac{\delta \tilde{E}}{\delta \m} }$ where $\tilde{E}$ is the micromagnetic energy density. eq~\eqref{Eq:LLG} can be derived from the Lagrangian, eq~\eqref{Def:Lag}, along with the generalized forces derived from the expression in eq~\eqref{Def:Force},
\begin{align}
	L &= \mu_0 M_s w \th \int dz \left( -\frac{m_z}{\gamma}\dot{\phi} - E[\m](z) \right) \label{Def:Lag} \\
	\delta W &= \mu_0 M_s w\th \int dz \left( -\frac{\alpha}{\gamma} \dot{\m} + \frac{\hbar}{e \mu_0 M_s \th} \left(\Is\times\m\right) \right).\delta\m \label{Def:Force}
\end{align}
where $\mu_0 M_s E = \tilde{E}$. In this section, we derive this equivalence. We take $\{m_z(z),\phi(z)\}$ as coordinates. Hence, we can write,
\begin{align}
	\delta\m &= \sin\theta\delta\phi \uvv{\Phi} + \delta \theta \uvv{\Theta}
\end{align}
where $\cos\theta = m_z$, $\uvv{\Phi}=(-\sin\phi,\cos\phi,0)$ and $\uvv{\Theta} = (\cos\theta\cos\phi, \cos\theta\sin\phi, -\sin\theta)$. This gives $\dot{\m}.\delta\m = \sin^2\theta \dot{\phi}\delta\phi + \dot{\theta}\delta\theta$ which translates to,
\begin{equation}
	\dot{\m}.\delta\m = \sin^2\theta \dot{\phi}\delta\phi + \frac{\dot{m}_z\delta m_z}{1-m_z^2} 
\end{equation}

One can notice that $\{ \m,\uvv{\Theta},\uvv{\Phi} \}$ forms an orthonormal right handed system. They are just the unit vectors in polar coordinates with $\m$ acting like the radius vector. Using Euler Lagrange equations by varying the action with respect to the coordinates, we get two equations of motion,
\begin{align}
	\frac{d}{dt} \frac{\delta L}{\delta \dot{\phi}} - \frac{\delta L}{\delta \phi} &= \frac{\delta W}{\delta \phi} \Rightarrow \nonumber \\
	-\dot{m}_z + \gamma \frac{\delta E}{\delta \phi} &= -\alpha (1-m_z^2) \dot{\phi} - \frac{\gamma \hbar}{e\mu_0M_s \th} \sqrt{1-m_z^2} \Is.\uvv{\Theta} \label{Eq:LLG1} \\
	\frac{d}{dt} \frac{\delta L}{\delta \dot{m}_z} - \frac{\delta L}{\delta m_z} &= \frac{\delta W}{\delta m_z} \Rightarrow \nonumber \\
	\dot{\phi} + \gamma\frac{\delta E}{\delta m_z} &= -\alpha \frac{\dot{m}_z}{1-m_z^2} - \frac{\gamma \hbar}{e\mu_0M_s \th} \frac{\Is.\uvv{\Phi}}{\sqrt{1-m_z^2}} \label{Eq:LLG2} 
\end{align}

The above equations are equivalent to the Landau-Lifshitz Gilbert (LLG) equation. To show this, we explicitly reduce down the terms in eq~\eqref{Eq:LLG1} to relevant terms in eq~\eqref{Eq:LLG}.
\begin{align}
	\gamma \frac{\delta E}{\delta \phi} &= \gamma \frac{\delta E}{\delta m_y}\sin\theta\cos\phi - \gamma \frac{\delta E}{\delta m_x}\sin\theta\sin\phi \\
&= -\gamma \left( \m\times\vct{H}_{\text{eff}} \right)_z \\
	\alpha (1-m_z^2)\dot{\phi} &= \alpha (1-m_z^2) \frac{m_x\dot{m}_y - m_y\dot{m}_x}{m_x^2 + m_y^2} \\
&= \alpha \left( \m\times\dot{\m} \right)_z \\
	\sqrt{1-m_z^2} \Is.\uvv{\Theta} &= \sin\theta\Is.\left(\m\times\uvv{\Phi}\right) \\
&= \Is.\left(\m\times(\uvv{z}\times\m)\right) \\
	&= -\left( \m\times(\Is\times\m) \right)_z 
\end{align}

From the above set of equalities, we can readily write eq~\eqref{Eq:LLG1} as,
\begin{equation} \label{Eq:LLG1:Dis}
	\frac{d \m_z}{dt} = -\gamma \left(\m\times\vct{H}_{\text{eff}}\right)_z + \alpha\left(\m\times\frac{d\m}{dt}\right)_z - \frac{\gamma \hbar}{e\mu_0 M_s \th}\left( \m\times\left(\Is(z)\times\m\right) \right)_z
\end{equation}
which is just the $z$-component of eq~\eqref{Eq:LLG}. Hence, one of the equations we get from Lagrangian is compatible with LLG. For the next equation, eq~\eqref{Eq:LLG2}, we proceed from eq~\eqref{Eq:LLG} and show that we can derive eq~\eqref{Eq:LLG2},
\begin{align}
	\uvv{\Phi}.\dot{\m} &= -\gamma \uvv{\Phi}.\left(\m\times\vct{H}_{\text{eff}}\right) + \alpha\uvv{\Phi}.\left(\m\times\dot{\m}\right) - \frac{\gamma\hbar}{e\mu_0M_s \th} \Is.\uvv{\Phi} \Rightarrow \nonumber \\
	\sin\theta \dot{\phi} &= -\gamma \uvv{\Theta}.\vct{H}_{\text{eff}} + \alpha \uvv{\Theta}.\dot{\m} - \frac{\gamma\hbar}{e\mu_0M_s \th} \Is.\uvv{\Phi} \Rightarrow \nonumber \\
	\sin\theta \dot{\phi} &= \gamma \left[ H_{\text{eff},z}\sin\theta - \cos\theta\left( H_{\text{eff},x}\cos\phi + H_{\text{eff},y}\sin\phi \right) \right] - \alpha \frac{\dot{m}_z}{\sin\theta} - \frac{\gamma\hbar}{e\mu_0M_s \th} \Is.\uvv{\Phi} \Rightarrow \nonumber \\
	\dot{\phi} &= \gamma \left[ H_{\text{eff},z} - \frac{m_z}{1-m_z^2} \left( H_{\text{eff},x}\cos\phi + H_{\text{eff},y}\sin\phi \right) \right] - \alpha \frac{\dot{m}_z}{1-m_z^2} - \frac{\gamma\hbar}{e\mu_0M_s \th} \frac{\Is.\uvv{\Phi}}{\sqrt{1-m_z^2}} \Rightarrow \nonumber \\
	\dot{\phi} &= -\gamma \frac{\delta E}{\delta m_z} - \alpha \frac{\dot{m}_z}{1-m_z^2} - \frac{\gamma \hbar}{e\mu_0M_s \th} \frac{\Is.\uvv{\Phi}}{\sqrt{1-m_z^2}}
\end{align}

Hence, we show that both eq~\eqref{Eq:LLG1} and eq~\eqref{Eq:LLG2} are satisfied if the original LLG equation eq~\eqref{Eq:LLG} is. As the number of independent equations in LLG is 2, we can say that the former two equations are equivalent to LLG.

\section{Derivation of the Rigid Domain Wall Approximation}

We concluded the last section by deriving a Lagrangian for the magnetization dynamics. In this section, we use rigid domain wall approximation to derive the equation of motion for the collective coordinates. This amounts to putting the following ansatz in the Lagrangian,
\begin{align} 
	\phi(z,t) &= \psi(t) \label{Exp:Ansatz:Phi} \\
	\theta(z,t) &= 2\tan^{-1}\left( e^{\frac{z-Z(t)}{\lambda(t)}} \right) \label{Exp:Ansatz:Theta} 
\end{align}

After we put this ansatz in the Lagrangian eq~\eqref{Def:Lag}, we can carry out the integration with respect to the $z$ co-ordinate. The energy functional (divided by $\mu_0M_s$) is,
\begin{equation} \label{Def:Ener} 
	E = \frac{\Aex}{\mu_0M_s} \left(\partial_z\m\right)^2 + \frac{\Hper}{2} m_y^2 - \frac{\Hpar}{2}m_z^2 - \m.\Hext
\end{equation}

Then, we will be left with a Lagrangian with $Z$, $\psi$ and $\lambda$ as coordinates. It is this final Lagrangian that will give us the equation of motion via the Euler-Lagrange equations. The generalized force due to STT, however, will have to be changed as well. By the prescription of work-energy, we have
\begin{equation}  \int dz \left(\Is\times\m\right).\delta\m = \Fp\delta\psi + \FZ\delta Z + \Fl\delta\lambda
\end{equation}
where $\{ \Fp, \FZ, \Fl \}$ are functions of $\{ Z,\psi,\lambda\}$ which we are going to find. For convenience in notation, we resolve $\Is$ as,
\begin{equation} \label{Def:Cur:Components} 
	\Is(z) = J_1(z) \m(z) + J_2(z)\m'(z) + J_3(z) \m(z)\times\m'(z)
\end{equation}
where $\m'(z)$ denotes the derivative of $\m(z)$ w.r.t $z$. The above expansion is valid only if $\m'$ is non-zero everywhere. This is true in the particular ansatz we are considering. We write down the relevant expressions now,
\begin{align}
	\m'.\left(\m\times\delta\m\right) &= -\sin^2\theta \frac{\delta\psi}{\lambda} \\
	\m'.\delta\m &= -\sin^2\theta\left( \frac{\delta Z}{\lambda^2} + \frac{\delta\lambda(z-Z)}{\lambda^3} \right)
\end{align}

Now, using the above equations, the term under consideration is (refer~\eqref{Def:Cur:Components} for definition of $J_i$),
\begin{align}
	\Is.\left(\m\times\delta\m\right)	&= J_2 \m'.(\m\times\delta\m) + J_3\m'.\delta\m \\
	&= -J_2 \theta' \sin\theta\delta\psi + J_3\theta'\delta\theta \\
&= -\sin^2(\theta(z)) \left[ J_2(z) \frac{\delta\psi(t)}{\lambda(t)} + J_3(z)\left( \frac{\delta Z}{\lambda^2} + \frac{\delta\lambda(z-Z)}{\lambda^3} \right) \right] \\
	\int dz \Is.\left(\m\times\delta \m\right) &= \Fp\delta{\psi} + \FZ\delta Z + \Fl\delta\lambda \\
	\Fp &= -\int \frac{dz}{\lambda} \frac{4}{\left[\exp\left(\frac{z-Z}{\lambda}\right) + \exp\left(-\frac{z-Z}{\lambda}\right)\right]^2} J_2(z) \label{Def:Fpsi} \\
	\FZ &= -\int \frac{dz}{\lambda^2} \frac{4}{\left[\exp\left(\frac{z-Z}{\lambda}\right) + \exp\left(-\frac{z-Z}{\lambda}\right)\right]^2} J_3(z) \label{Def:FZ} \\
	\Fl &= -\int \frac{dz}{\lambda^3} \frac{4(z-Z)}{\left[\exp\left(\frac{z-Z}{\lambda}\right) + \exp\left(-\frac{z-Z}{\lambda}\right)\right]^2} J_3(z) \label{Def:Flambda}
\end{align}
So far, it was generally applicable for any distribution of spin currents. For the case under consideration, we focus on a specific spin current distribution such that $J^z$ is a constant, say $J^s_z$, for $z<0$ and zero for $z>0$. $J^x$ and $J^y$ are zero everywhere. Then, we get $\FZ=\Fl=0$ with $\Fp$ given by,
\begin{equation}
	\Fp = \frac{2 J^s_z \lambda}{1+\exp\left(\frac{2Z}{\lambda}\right)}
\end{equation}

Now, we calculate the integrals with respect to the $z$ co-ordinate for the terms in the main Lagrangian. We list down the equality among integrals,
\begin{align}
	\int dz	\frac{-m_z}{\gamma} \dot{\phi} &= \frac{-2Z}{\gamma}\dot{\psi} \\
	\int dz \frac{\Aex}{\mu_0 M_s} (\partial_x\m)^2 &= \frac{2\Aex}{\mu_0 M_s\lambda} \\
	\int dz	\frac{\Hper}{2} (\m.\uvv{y})^2 &= \Hper\lambda\sin^2\psi \\
	\int dz	\frac{-\Hpar}{2}(\m.\uvv{z})^2 &= \Hpar (l-\lambda) \\
	\int dz	\frac{\alpha}{\gamma}\dot{\m}.\delta\m &= \frac{2\alpha\lambda}{\gamma} \left( \frac{\dot{Z}\delta Z}{\lambda^2} + \dot{\psi} \delta\psi \right) + \frac{\alpha\pi^2}{6\gamma}\frac{\dot{\lambda}\delta\lambda}{\lambda} \\
	\int dz \left(-\m.\vct{H}_{\text{ext}}\right) &= -2 H_{\text{ext}} Z
\end{align}

The final Lagrangian along with work is given by,
\begin{align}
	L &= \mu_0 M_s w\th \left[ \frac{-2Z}{\gamma} \dot{\psi} - \frac{2\Aex}{\mu_0M_s\lambda} - \Hper\lambda\sin^2\psi + \Hpar(l-\lambda) + \pi\lambda \left(H_x\cos\psi + H_y\sin\psi \right) + 2 H_zZ \right] \\
	\delta W &= -\mu_0 M_s w\th \left[ \frac{2\alpha\lambda}{\gamma} \left( \frac{\dot{Z}.\delta Z}{\lambda^2} + \dot{\psi}.\delta\psi \right) + \frac{\alpha\pi^2}{6\gamma}\frac{\dot{\lambda}.\delta\lambda}{\lambda} \right] + \FZ\delta Z + \Fp\delta\psi + \Fl\delta\lambda 
\end{align}

The Euler-Lagrange equations then lead to,
\begin{align}
	-\dot{\psi} + \gamma H_{\text{ext}} &= \frac{\alpha\dot{Z}}{\lambda} \label{Eq:DWNoiseless:Psi} \\
	\frac{\dot{Z}}{\lambda} - \gamma\Hper\sin\psi\cos\psi &= \alpha\dot{\psi} - \frac{\gamma \hbar J^s_z}{e \mu_0 M_s \th} \frac{1}{1+\exp\left(\frac{2Z}{\lambda}\right)} \label{Eq:DWNoiseless:Z} \\
	\frac{2\Aex}{\mu_0M_s\lambda^2} - \Hper\sin^2\psi - \Hpar &= \frac{\alpha\pi^2}{6\gamma} \frac{\dot{\lambda}}{\lambda} \label{Eq:DWNoiseless:Lambda}
\end{align}
Note that we have applied a field in $-z$ direction in the paper and hence in the notation of the paper, $H_{\text{ext}} = -H$. Similarly, the sign of $J^s_z$ here is opposite of that in the paper.

So far, we have treated all coordinates $\{Z, \psi, \lambda\}$ on equal footing. However, $\lambda$ is different from the other two in the sense that it changes the shape of Domain Wall instead of just coherent translation ($Z$) and rotation ($\psi$). Hence, under the validity of the rigid domain wall, the variation in $\lambda$ should be small. This can be verified through numerical simulations as well. Using this approximation as a guide, we neglect the time derivative term (which is also proportional to $\alpha$, thereby helping the approximation) in eq~\eqref{Eq:DWNoiseless:Lambda}. Hence, we arrive at the equation of motion in $\{ Z, \psi \}$ with $\lambda$ as a function of $\psi$ instead of an independent variable, given by ${\displaystyle \lambda(\psi) = \sqrt{\frac{2\Aex}{\mu_0 M_s (\Hpar + \Hper\sin^2\psi)}} }$.

\section{Noise Analysis}
At $T=0$, rigid domain wall approximation correspond to eqs~\eqref{Eq:DWNoiseless:Psi} and \eqref{Eq:DWNoiseless:Z} with qualifications about $\lambda$ from the subsequent paragraph. At non-zero temperature, we have to add terms pertaining to noise in the aformentioned equations. In this section, we derive these terms and discuss the simulations of the stochastic differential equations. For analytical simplicity, we only consider the case when $\Hper\ll\Hpar$, such that $\lambda$ can be taken to be a constant. Also, we consider the case of equilibrium without an external magnetic field or spin current and then assume that the noise sources remain the same in non-equilibrium. Under such circumstances, we can rewrite the final set of equations more generally by the following definitions,
\begin{align}
	p &= \sqrt{\frac{2\mu_0M_sw\th}{\gamma}}\psi \label{Def:pNoise} \\ 
	x &= \sqrt{\frac{2\mu_0M_sw\th}{\gamma}} Z \label{Def:xNoise} \\
	\En &= \mu_0 M_s \Hper w\th\lambda\sin^2\psi \label{Def:EnNoise}
\end{align}
Note that with $\lambda$ as a constant, the only term remaining in energy is the hard axis anisotropy. After some calculations, we can reduce eqs~\eqref{Eq:DWNoiseless:Psi} and \eqref{Eq:DWNoiseless:Z} in terms of $x$ and $p$ as,
\begin{align}
	\dot{x} &= \frac{1}{1+\alpha^2} \left( \frac{\partial \En}{\partial p} - \alpha\lambda\frac{\partial \En}{\partial x} \right) + g_x N_x \\
	\dot{p} &= -\frac{1}{1+\alpha^2} \left(\frac{\partial \En}{\partial x} + \frac{\alpha}{\lambda} \frac{\partial \En}{\partial p} \right) + g_p N_p
\end{align}
where we have added two uncorrelated white source, $N_x$ and $N_p$; $g_x$ and $g_p$ are assumed to be independent of $Z$ and $\psi$. We assume ${\displaystyle \left\langle N(t) N(t') \right\rangle = \frac{2\alpha k_B T}{1+\alpha^2} \delta(t-t')}$ for both $N_x$ and $N_p$. These assumptions are justified by deriving the expressions of $g_x$ and $g_p$ from the Fokker-Planck equations corresponding to the above Langevin equations. They are given as,
\begin{align}
	\frac{\partial P}{\partial t} &= -\frac{\partial}{\partial x} \left[D_x^{(1)} P\right] - \frac{\partial}{\partial p} \left[D_p^{(1)} P\right] + \frac{\partial^2 }{\partial x^2} \left[D_x^{(2)} P\right] + \frac{\partial^2 }{\partial p^2} \left[D_p^{(2)} P\right]
\end{align}
where the drift and diffusion coefficients are given by,
\begin{align}
	D_x^{(1)} &= \frac{1}{1+\alpha^2} \left( \frac{\partial \En}{\partial p} - \alpha\lambda\frac{\partial \En}{\partial x} \right) & D_p^{(1)} &= -\frac{1}{1+\alpha^2} \left(\frac{\partial \En}{\partial x} + \frac{\alpha}{\lambda} \frac{\partial \En}{\partial p} \right) \\
	D_x^{(2)} &= \alpha k_BT g_x^2 & D_p^{(2)} &= \alpha k_BT g_p^2
\end{align}

After a series of elementary manipulations, we can get the expressions for $g_x^2 = \lambda$ and $g_p^2 = 1/\lambda$ by demanding that $P=C\exp(-\En/k_BT)$ satisfies the Fokker-Planck equations in equilibrium, where $C$ is some constant. Now we revert back to the original coordinates, $\{Z,\psi\}$. We define two noise terms, $N_Z = N_x-\alpha N_p$ and $N_{\psi} = \alpha N_x + N_p$ to simplify the final expression as,
\begin{align}
	\frac{\dot{Z}}{\lambda} &= \gamma \Hper\sin\psi\cos\psi + \alpha \dot{\psi} + \frac{\gamma\hbar J^s_z}{e\mu_0M_s\th}\frac{1}{1+\exp(2Z/\lambda)} + \sqrt{\frac{\gamma}{2\mu_0M_s w\th \lambda} } N_Z \label{Eq:Noisy:Z} \\
	\dot{\psi} &= \gamma H_{\text{ext}} - \frac{\alpha\dot{Z}}{\lambda} + \sqrt{\frac{\gamma}{2\mu_0M_s w\th \lambda}} N_{\psi} \label{Eq:Noisy:Psi}
\end{align}
where we have added the external magnetic field and spin current back. Note that both $N_Z$ and $N_{\psi}$ are uncorrelated white noise sources with $\left\langle N(t) N(t') \right\rangle = 2\alpha k_BT\delta(t-t')$.

We then numerically simulated eqs~\eqref{Eq:Noisy:Z} and \eqref{Eq:Noisy:Psi} using Stratonovich calculus and second order Heunn scheme. We wrote the code in MATLAB by discretising time in units of $1ps$ and assuming the parameters as $M_s=800kA/m$, $\Aex = 13 pJ/m$, $\Hpar=32.9 kA/m$, $\Hper = 16.6 kA/m$, $\gamma=2.21\times10^5 m/(A-s)$ and $\alpha = 0.007$. To verify the code under equilibrium, we added a pinning potential of the form $0.5 k_{\text{Pin}} Z^2$, with $k_{\text{Pin}}=15.12\mu J/m^3$, and assumed a low temperature of $50K$ such that $\sin^2\psi \approx \psi^2$. Under such circumstances, the energy (refer eq~\eqref{Def:EnNoise}) becomes analogous to a particle in a harmonic potential and we verified that the equipartition theorem holds by finding the average value of $Z^2$ and $\psi^2$. All averages as well as any other statistical values were calculated after removing the initial motion for $40ns$ to allow the system to achieve time invariance.

We then applied the external inputs and assumed a room temperature of $300 K$. We simulated the motion for $40\mu s$ and found the spectral function (again after removing the initial $40ns$ of motion). The simulation was run for various values of $H_{\text{ext}}$ and $J^s_z$ and we extracted the Q-factor in all the cases, which turned out to be between $500$ and $1500$. The simulation result for three cases is shown in the paper (see Fig 4 of the paper).

\section{Micromagnetic Simulations}

The theoretical analysis in the main paper was carried out using rigid domain wall approximation imposed on a 1 dimensional LLG with effective parameters for anisotropy. For a more quantitative treatment, we must take into account the dipolar interaction precisely instead of masking it with a local effective field as in eq~\eqref{Def:Ener}. This, in general, might be a difficult problem to solve analytically. Hence, we simulate a magnetic domain wall under the conditions given in the paper to verify if the oscillations are happening and compare the motion with the one derived from theoretical analysis.

We consider a magnetic thin film of thickness $3nm$ (along $y$-axis), width $100nm$ (along $x$-axis) and length $800nm$ (along $z$-axis). Consider the LLG equation given in eq~\eqref{Eq:LLG} with the effective magnetic field given by a sum of exchange field, crystalline anisotropy, field due to other dipoles and an external magnetic field.
\begin{equation}
	\vct{H}_{\text{eff}} = \frac{2\Aex}{\mu_0 M_s} \partial^2_z\m + \frac{2K_a}{\mu_0M_s}m_y\uvv{y} + \Hext + \vct{H}_{\text{dip}}
\end{equation}
where we assume crystalline anisotropy, $K_a$, to reduce the hard axis anisotropy due to demagnetisation field (this is required for rigid domain wall approximation to work); $\vct{H}_{\text{dip}}$ is the effective field due to the dipolar interaction between magnetizations. $\vct{H}_{\text{dip}}$ has been calculated as discussed by Miltat et al.\cite{Miltat}

\begin{figure}
	\centering \includegraphics[width=.5\textwidth,keepaspectratio]{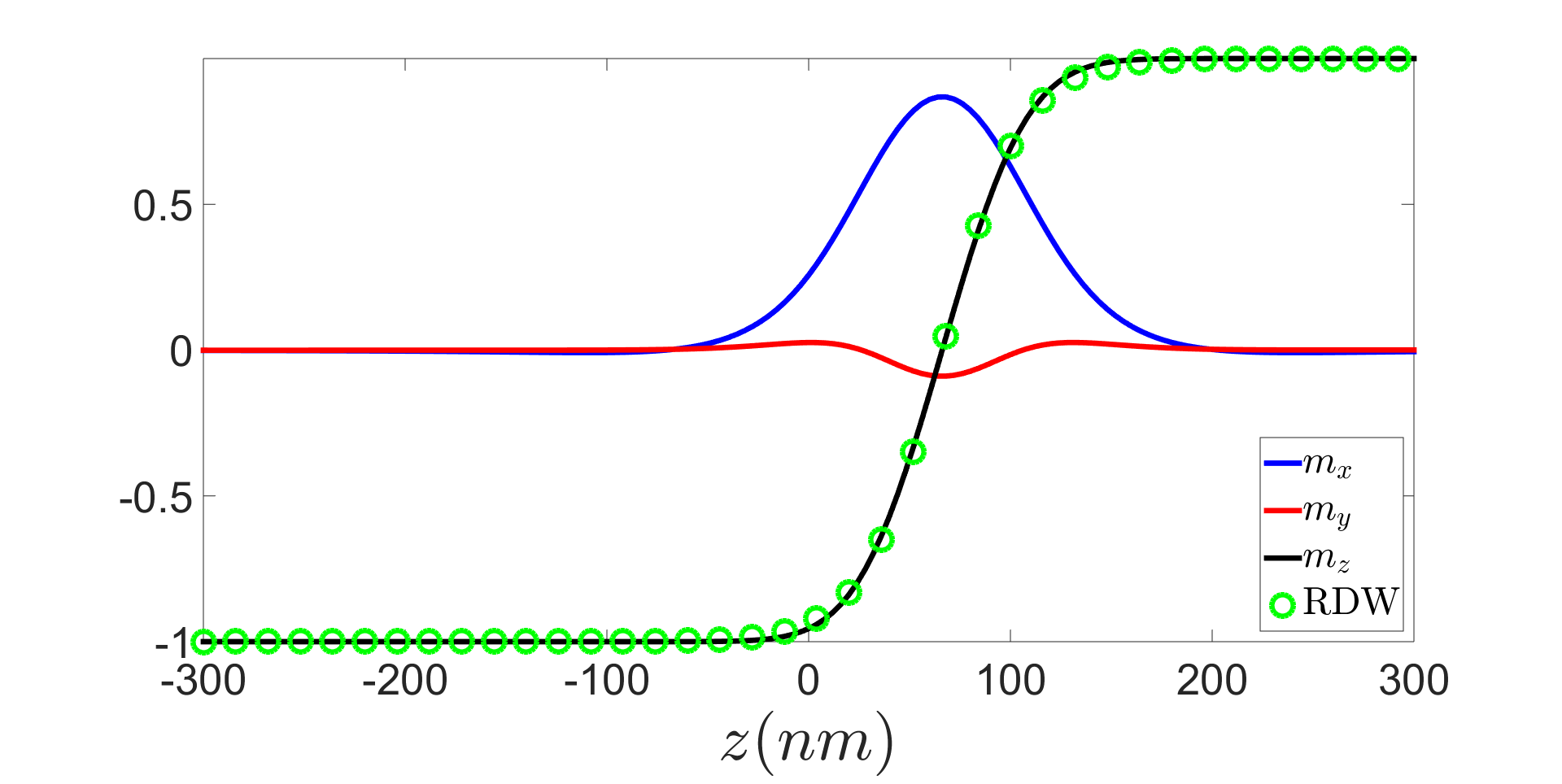}
	\caption{The equilibrium magnetization along the length of the magnet. Here $\{m_x,m_y,m_z\}$ denotes the average of the quantities taken along the width, $x$-axis. The dots (`RDW') refers to the fitted waveform of $m_z$ in rigid domain wall approximation.} \label{Fig:Eq}
\end{figure}

We use finite difference method and fourth order Runge Kutta with a fixed time step of $0.2ps$ to simulate a tail-to-tail domain wall. As the thickness of the film is small compared to typical exchange length, we perform a 2-dimensional simulation with discretization in blocks of $4nm\times4nm$. The values of parameters are taken to be $\Aex=13 pJ/m$, $M_s = 800 kA/m$, $K_a = 0.35 MJ/m^3$ and $\alpha = 0.007$. To achieve the domain wall configuration, we simulate an infinite strip by adding additional magnetic charges at the two edges along the length\cite{InfDW}. These magnetic charges are due to the remaning left and right parts of the infinite strip, which are assumed to have a constant magnetization pointing along $\pm z$-axis. Under such conditions, we let the system equilibriate for a sufficient amount of time. After equilibrium, we observe that $m_y$ is zero in the bulk and develop small non-zero values at the edges along the width. We average $\{m_x,m_y,m_z\}$ along the strip width ($x$-axis) to get the plots given in Fig~\ref{Fig:Eq}. We have also shown the theoretical waveform derived from rigid domain wall approximation, ${\displaystyle m_z(z) = \tanh\left(\frac{z-Z}{\lambda}\right)}$ with the domain wall position, $Z$ and width, $\lambda$ chosen to fit the micromagnetic result. The system equilibriates at $Z=64.2nm$ and $\lambda=38.9nm$. We note that the initial location of the domain wall, $Z$, is dependent on the initial conditions of the simulations while the width, $\lambda$, is independent of it.

\begin{figure}
	\centering \includegraphics[width=.5\textwidth,keepaspectratio]{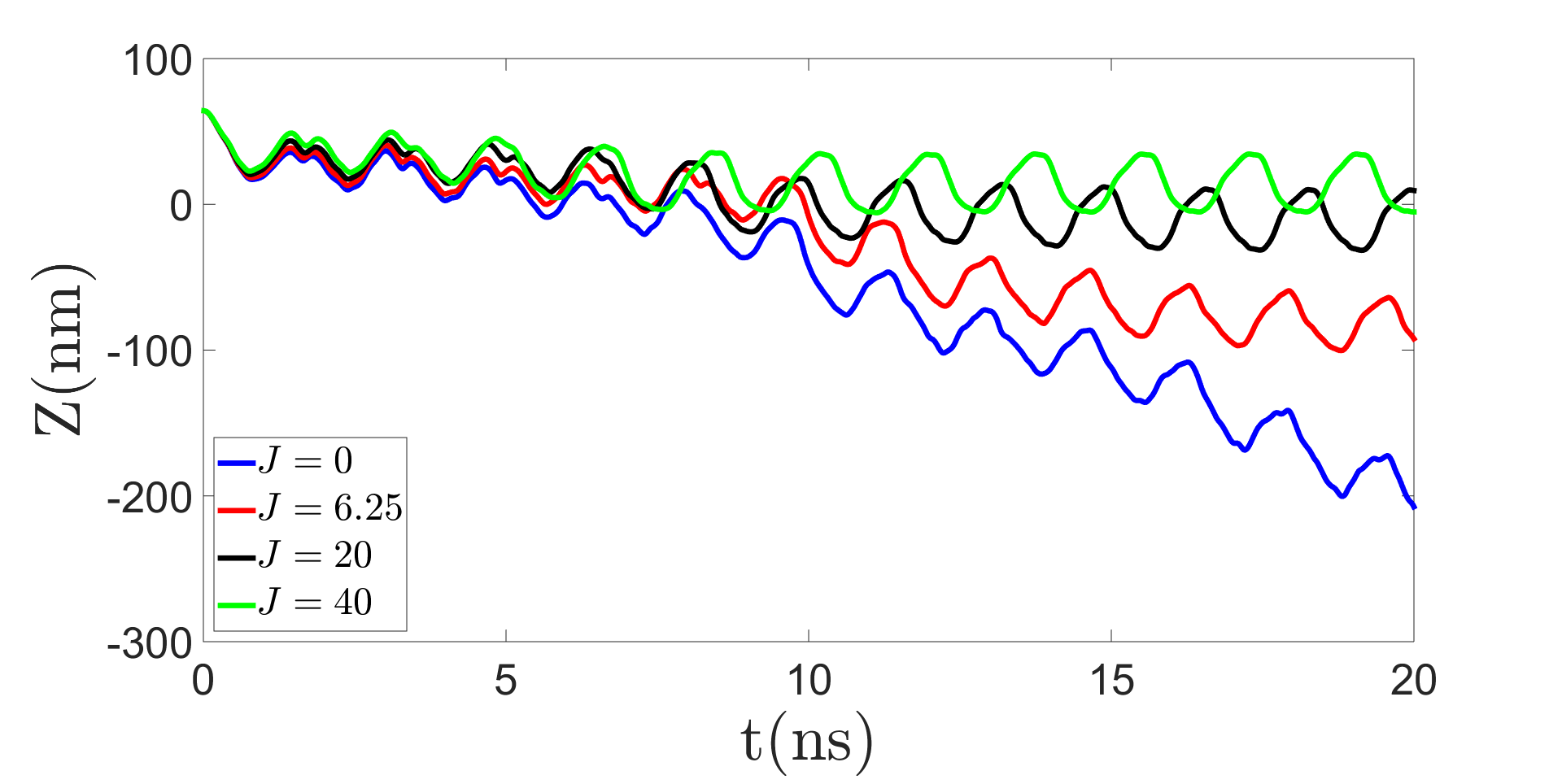}
	\caption{The motion of the domain wall under various values of spin current density with a magnetic field of $8.75 kA/m$. The spin current is restricted to be in one half of the magnet as discussed in the paper.} \label{Fig:VarCur}
\end{figure}

We then apply an external magnetic field of $8.75 kA/m$ and vary the vertical spin current density to see the motion under these stimuli. The spin current applied is restricted to one half of the magnet as discussed in the paper. Fig~\ref{Fig:VarCur} shows the motion under various values of spin current density, all in $GA/m^2$. The domain wall position has been defined as $(-l/2)*\langle m_z \rangle_x$ where the average is taken along $x$-direction. Note that we could as well define $Z$ to be the point where $\langle m_z \rangle_x$ is zero. However, this distinction doesn't quantitatively affect the dynamics. At $J=0$, we get back the usual oscillatory and drifting motion of the domain wall position. For non-zero small $J$, we still have a drifting domain wall but with a smaller velocity. For higher $J$, we get pure oscillations with the average domain wall position dependent on the spin current as discussed in the paper.

\begin{figure}
	\centering \includegraphics[width=.5\textwidth,keepaspectratio]{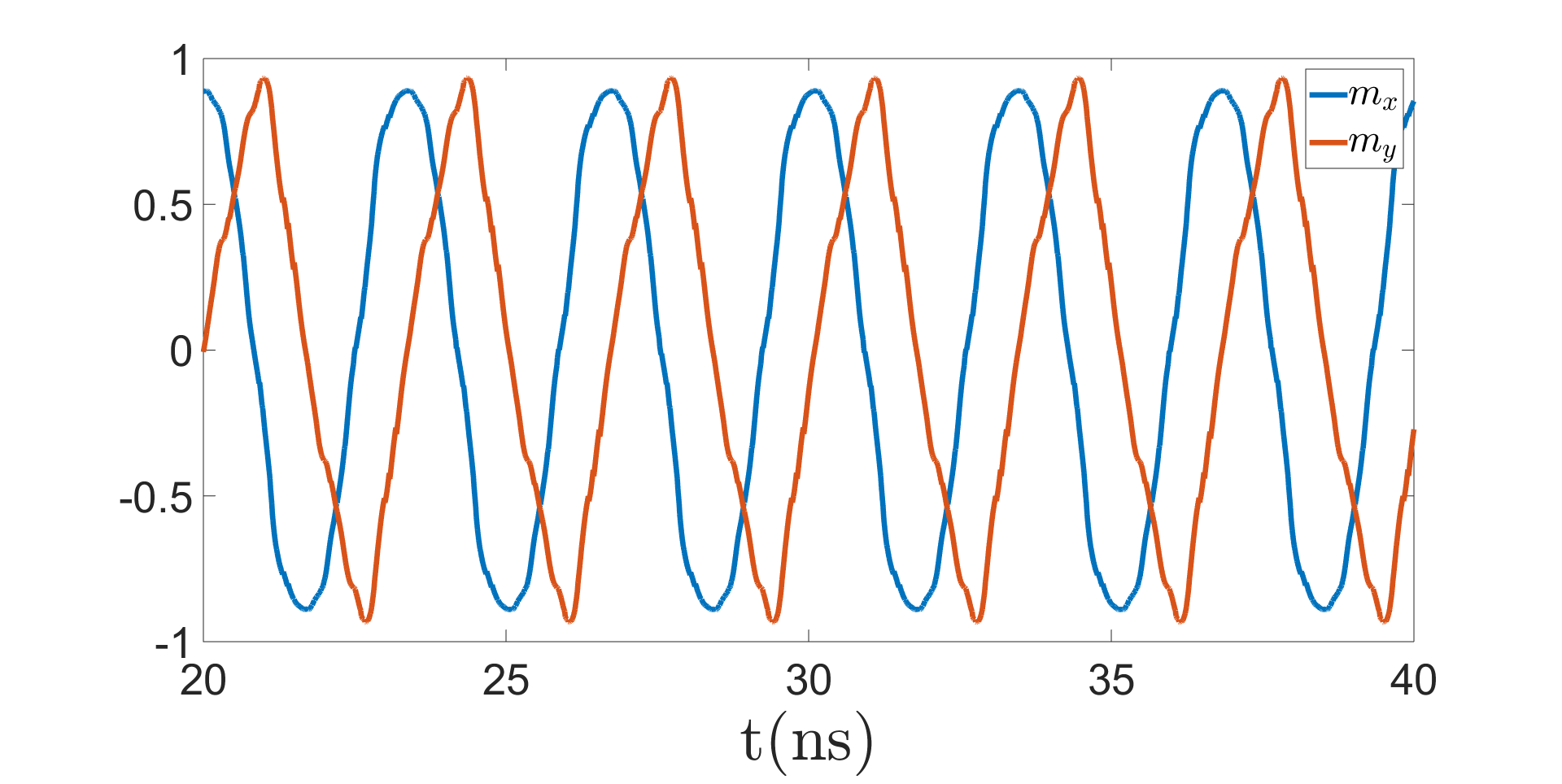}
	\caption{Depicting rotation in domain wall. Here $m_y$ means the average value of $m_y$ at the instantaneous position of the domain wall. The averaging is done across the width. $m_x$ is computed similarly.} \label{Fig:Rot}
\end{figure}

\begin{figure}
	\centering \includegraphics[width=.5\textwidth,keepaspectratio]{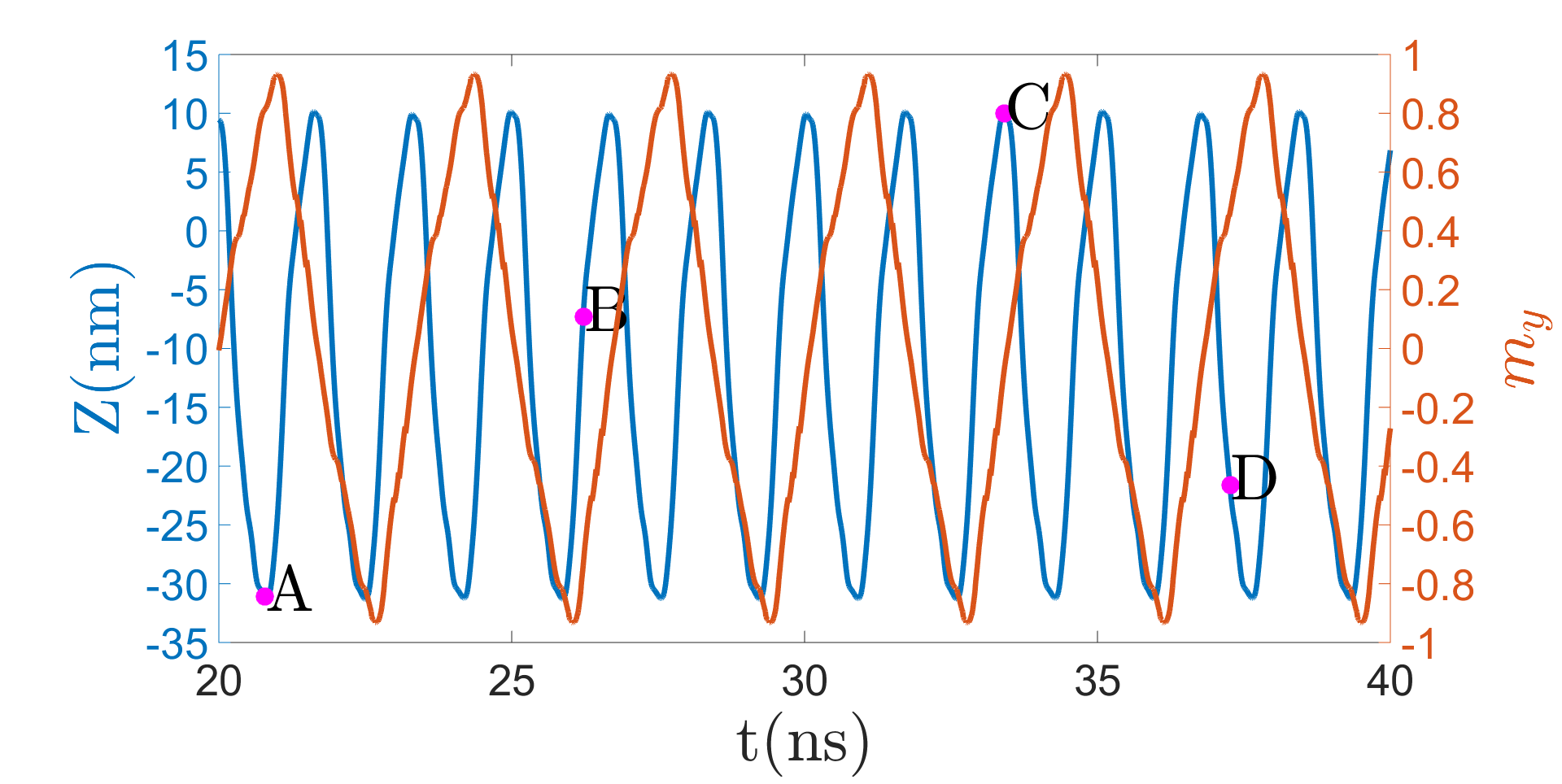}
	\caption{Illustrating the phase relation between the domain wall position and its rotation. Here again, $m_y$ denotes the average value across the width at the instantaneous position of the domain wall.} \label{Fig:OscPhase}
\end{figure}

Now, we focus on the case of $J=20 GA/m^2$ where oscillations are stable. The oscillation frequency is $0.56 GHz$ which agrees with the one derived from the rigid domain wall approximation in the paper, $0.61 GHz$. As discussed in the paper, we have both translatory oscillations as governed by $Z$ and rotation of the wall as governed by $\psi$. To show the rotation we plot the time evolution of $\langle m_y \rangle_x$ and $\langle m_x \rangle_x$ at the instantaneous domain wall position in Fig~\ref{Fig:Rot}. We can see that they are oscillating with a phase difference of close to $\pi/2$. In rigid domain wall approximation, the oscillations of $\psi$ occur with half the frequency of $Z$. This can be explicitly noted in micromagnetic simulations as well by plotting $\langle m_y \rangle_x$ at $z=Z(t)$ as a function of time and comparing it to $Z(t)$ itself as shown in Fig~\ref{Fig:OscPhase}. From eq~\eqref{Eq:DWNoiseless:Z}, we can see if the domain wall is sufficiently outside the region of non-zero spin current, $Z\gg\lambda$, the velocity of the domain wall is zero at $\psi = 0,\pm\pi/2,\pi$. Even though this condition is never achieved in the shown oscillations, it can be approximately noticed from the plot at the maxima of the domain wall position (where $Z$ is still less than $\lambda$).

\begin{figure}
	\centering \includegraphics[width=.5\textwidth,keepaspectratio]{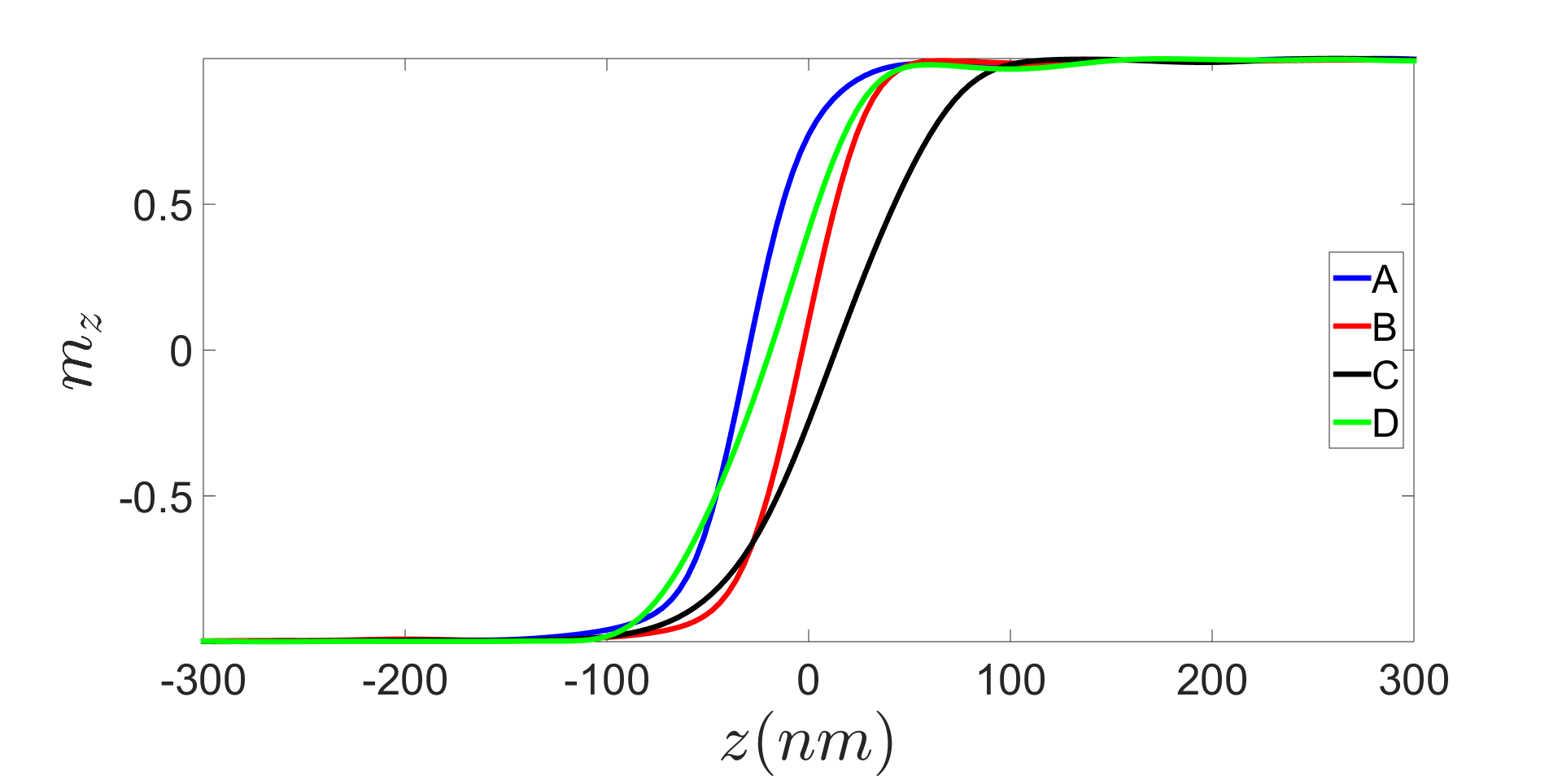}
	\caption{The plot of $m_z$ averaged across the width for four different instances in the oscillations. This depicts the variation in width during different phases of the oscillations.} \label{Fig:ExpandOsc}
\end{figure}

Fig~\ref{Fig:ExpandOsc} is a plot of $\langle m_z \rangle_x$ at the four points marked as $\{A,B,C,D\}$ in Fig~\ref{Fig:OscPhase}. These waveforms can be fitted with rigid domain wall approximation with the domain wall position and width as fitting parameters. Using this, we can verify that at least for the averaged value of $m_z$, the rigid domain wall approximation is valid throughout the oscillations. This also illustrates the oscillations of the width of the domain wall.

To compare the micromagnetic simulations with theoretical analysis, we need to define $\Hper$ and $\Hpar$ through micromagnetics. The definition is given in terms of mean field of a mono domain magnet. We extract $H_x$ as the mean field when all the magnetic moments are pointing along $x$ axis. $H_y$ and $H_z$ are defined analogously. We then define $\Hpar$ to be $H_x-H_z$ and $\Hper$ as $H_y-H_x$. We note however that defining a particular $\Hper$ and $\Hpar$ can only give an estimate of the observables and cannot be used for quantitative correctness\cite{MM_RDW}. We get the values as $\Hpar \approx 32.9 kA/m$ and $\Hper \approx 16.6 kA/m$. Using these as effective parameters in the LLG equation, we calculate the amplitude of oscillations through the rigid domain wall approximation. We find that the amplitude is of the same order as the micromagnetic prediction. The amplitude in micromagnetic simulations is approximately twice of the one derived from theoretical analysis in the paper. The factor of 2 could be because of a specific definition of $\Hper$ and $\Hpar$ and hence cannot be predicted easily. This is an evidence of the validity of the theoretical analysis carried out in the main paper. 

\bibliography{References}